\documentclass[twocolumn,prx,aps,superscriptaddress,nofootinbib]{revtex4}
\usepackage{graphicx,dcolumn,bm,color,framed}
\usepackage[utf8]{inputenc}

\usepackage{amssymb}
\usepackage{graphicx}
\usepackage{epsf,epsfig}
\usepackage{bm}
\usepackage{bbm}
\usepackage{amsfonts}
\usepackage{amsmath}
\usepackage{graphics}
\usepackage[normalem]{ulem}
\usepackage{color}

\begin{document}

\title{The role of electron-electron collisions for magnetotransport at intermediate temperatures}

\author{Woo-Ram Lee}
\affiliation{Department of Physics and Astronomy, The University of Alabama, Tuscaloosa, Alabama 35487, USA}
\affiliation{Department of Physics, Virginia Tech, Blacksburg, Virginia 24061, USA}
\author{Alexander M. Finkel'stein}
\affiliation{Department of Physics and Astronomy, Texas A\&M University, College Station, Texas 77843-4242, USA}
\affiliation{Department of Condensed Matter Physics, The Weizmann Institute of Science, Rehovot 76100, Israel}
\author{Georg Schwiete}
\affiliation{Department of Physics and Astronomy, The University of Alabama, Tuscaloosa, Alabama 35487, USA}

\begin{abstract}
We discuss galvanomagnetic and thermomagnetic effects in disordered electronic systems focusing on intermediate temperatures, for which electron-electron scattering and electron-impurity scattering occur at similar rates, while phonon-related effects can be neglected. In particular, we explore how electric and thermal currents driven either by an electric field or by a temperature gradient are affected by the interplay of momentum-dependent electron-impurity scattering, electron-electron scattering, and the presence of a magnetic field. We find that the electric resistance, the Seebeck coefficient and the Nernst coefficient are particularly sensitive to the momentum dependence of the electron-impurity scattering rate at intermediate temperatures. A sufficiently strong momentum dependence of the electron-impurity scattering rate can induce a sign change of the Seebeck coefficient. This sign change can be suppressed by a perpendicular magnetic field. The temperature and magnetic field dependence of the Seebeck coefficient can be used for measuring the magnitude of the electron-impurity and electron-electron scattering rates. The Nernst coefficient vanishes for momentum-independent electron-impurity scattering, but displays a maximum at finite temperatures once the momentum dependence is accounted for. By contrast, the Hall coefficient and the Righi-Leduc coefficient display only a weak dependence on the momentum dependence of the electron-impurity scattering at intermediate temperatures.
\end{abstract}

\maketitle

\section{Introduction}

Galvanomagnetic and thermomagnetic phenomena in metals and semiconductors have been discussed extensively in the literature when electron-phonon and electron-impurity collisions are the dominant scattering mechanisms \cite{Tsuji58,Beer63,Ziman01,Gantmakher12}. Indeed, elastic electron-impurity scattering events dominate at low temperatures, when phonons freeze out and inelastic scattering of electrons off each other becomes ineffective due to the Pauli-exclusion principle. At elevated temperatures, in turn, electron-phonon scattering is responsible for the leading temperature dependence of the transport coefficients. In between these two transport regimes, the influence of electron-electron collisions on the transport coefficients may become visible. If the corresponding interval of temperatures is sufficiently broad, then hydrodynamic behavior can be observed on its upper end, once electron-electron scattering occurs much more frequently than electron-impurity scattering \cite{Gurzhi63,deJong95,Andreev11,Mahajan13,Xie16,Moll16,Crossno16,Bandurin16,Narozhny17,Guo17,Levchenko17,Lucas18,Zarenia19}. Here, we are mainly interested in an intermediate temperature regime, for which these two scattering processes occur at similar rates \cite{Principi15,Lucas18,Lee20}.

Throughout the paper, we assume that Umklapp scattering is prohibited or ineffective. In this case, electron-electron collisions conserve the total momentum, but may still influence the electric conductivity indirectly via a redistribution of occupation numbers in momentum space. This redistribution reveals itself in the temperature dependence of the electric conductivity $\sigma$, for example, if the electron-impurity scattering is momentum dependent \cite{Keyes58,Gantmakher12}. The effect is well illustrated by the two differing results obtained for the conductivity in the electron-impurity and electron-electron scattering dominated transport regimes, where $\sigma\propto \langle \tau_{ei,\bf p}\rangle$, and $\sigma\propto \langle 1/\tau_{ei,\bf p}\rangle^{-1}$, respectively \cite{Keyes58}. Here, $\tau_{ei,{\bf p}}$ is the electron-impurity scattering rate, and the angular bracket symbolizes a (weighted) thermal average. The crossover between these two limits has also been explored~\cite{Keyes58}. The combined influence of electron-electron scattering and momentum-dependent electron-impurity scattering on other transport coefficients in the intermediate temperature regime is less understood. Recently, we discussed the thermal and thermoelectric transport coefficients in this context \cite{Lee20}. We found, in particular, that the Seebeck coefficient can develop a non-monotonic temperature dependence accompanied by a sign change under the influence of the two scattering processes. Here, we generalize these studies to include a magnetic field. The inclusion of a magnetic field also opens the way to discussing transverse effects, characterized by the Hall, Nernst and Righi-Leduc coefficients. In order to achieve this goal, we restrict ourselves to isotropic systems with a quadratic dispersion. We base our considerations on a Boltzmann equation approach in which both electron-impurity and electron-electron collision integrals are treated in the relaxation time approximation. This simple model allows us to find compact and transparent expressions for the different coefficients, and to highlight the influence of the momentum dependence of $\tau_{ei,{\bf p}}$.

The Seebeck coefficient displays a particularly interesting behavior at intermediate temperatures. We find that a weak perpendicular magnetic field reduces the maximum in the temperature dependence of this coefficient and stronger magnetic fields can even suppress the sign change, making this coefficient a monotonically decreasing function of temperature. The obtained temperature and magnetic field dependence of this coefficient closely resembles the behavior observed for the Seebeck coefficient of Si:P on the metallic side of the metal-insulator transition, Ref.~\cite{Lakner93}. We therefore devote special attention to this coefficient. The Nernst coefficient is also strongly affected by the momentum dependence of the electron-impurity scattering rate. It vanishes in the absence of this momentum dependence in our model, and shows a characteristic maximum determined by the competition between elastic and inelastic scattering if it is present.

The paper is organized as follows. In Sec.~\ref{sec:PhenEq}, we review the phenomenological equations defining the galvanomagnetic and thermomagnetic transport coefficients, as well as the relevant Onsager relations. In Sec.~\ref{sec:LinBoltz}, we introduce the Boltzmann equation and solve it to obtain the linearized distribution function. In Sec.~\ref{sec:TransCoeff}, we find the nine transport coefficients which characterize electric, thermal and thermoelectric transport in three dimensions for an arbitrary magnetic field direction and discuss the geometry of the transport processes. Sec.~\ref{sec:discussion} and the supplementary material \cite{Suppl} are devoted to a discussion of the temperature and magnetic field dependence of the transport coefficients, with particular emphasis on the role of the momentum dependence of $\tau_{ei,{\bf p}}$. We conclude in Sec.~\ref{sec:conclusion}. Appendix~\ref{app:derivation} contains some technical details of the calculation.

\section{Phenomenological relations}
\label{sec:PhenEq}
The linear response relation between the electric and thermal currents ${\bf J}_E$ and ${\bf J}_T$, respectively, and the electric field ${\bf E}$ or thermal gradient $\nabla T$ driving these currents can be parametrized as \cite{Electrochem}
\begin{align}
\left(\begin{array}{c} {\bf J}_E\\{\bf J}_T\end{array}\right)=\left(\begin{array}{cc} \hat{\sigma}&\hat{\mathcal{M}}\\\hat{\mathcal{N}}&\hat{\mathcal{L}}\end{array}\right)\left(\begin{array}{c} {\bf E}\\ {\nabla T}\end{array}\right).
\label{eq:linear_response}
\end{align}
The components of the coefficient matrix are the conductivity tensor $\hat{\sigma}$, the thermal flow tensor $\hat{\mathcal{L}}$, and the two cross effect tensors $\hat{\mathcal{N}}$ and $\hat{\mathcal{M}}$.

For comparison with experiment, it is often convenient to use the electric current as an independent variable instead of the electric field. To achieve this, one may resolve the equation for the electric current ${\bf J}_E$ encoded in Eq.~\eqref{eq:linear_response} for ${\bf E}$, resulting in ${\bf E}=\hat{\sigma}^{-1}{\bf J}_E-\hat{\sigma}^{-1}\hat{\mathcal{M}}\nabla T$. After entering with this result into the equation for ${\bf J}_T$, one finds a new pair of equations which can also be expressed in a matrix form as
\begin{align}
\left(\begin{array}{c} {\bf E}\\{\bf J}_T\end{array}\right)&=\left(\begin{array}{cc} \hat{\rho}&-\hat{\alpha}\\ \hat{\pi}&\hat{\kappa}\end{array}\right)\left(\begin{array}{cc}{\bf J}_E\\-\nabla T\end{array}\right).\label{eq:experiment}
\end{align}
Here, we introduced the resistivity tensor $\hat{\rho}$, the thermoelectric power tensor $\hat{\alpha}$, the Peltier coefficient tensor $\hat{\pi}$, and the thermal conductivity tensor $\hat{\kappa}$. The relation between the tensors used in Eq.~\eqref{eq:experiment} and those introduced in Eq.~\eqref{eq:linear_response} is as follows
\begin{align}
\hat{\rho}&=\hat{\sigma}^{-1},\qquad \hat{\alpha}=-\hat{\rho}\hat{\mathcal{M}}{\color{red},}
\\
\hat{\pi}&=\hat{\mathcal{N}}\hat{\rho},\qquad \hat{\kappa}=\hat{\mathcal{N}}\hat{\rho}\hat{\mathcal{M}}-\hat{\mathcal{L}}.
\end{align}
The linear response relation in Eq.~\eqref{eq:linear_response} is often formulated with the help of $\hat{\alpha}$, $\hat{\pi}$ and $\hat{\kappa}$ instead of $\hat{\mathcal{M}}$, $\hat{\mathcal{N}}$, and $\hat{\mathcal{L}}$, via the relations $\hat{\mathcal{M}}=-\hat{\sigma}\hat{\alpha}$, $\hat{\mathcal{N}}=\hat{\pi}\hat{\sigma}$, and $\hat{\mathcal{L}}=-(\hat{\pi}\hat{\sigma}\hat{\alpha}+\hat{\kappa})$.

The components of the tensors $\hat{\sigma}$, $\hat{\mathcal{M}}$, $\hat{\mathcal{N}}$, $\hat{\mathcal{L}}$ in Eq.~\eqref{eq:linear_response} and, respectively, $\hat{\rho}$, $\hat{\alpha}$, $\hat{\pi}$, $\hat{\kappa}$ in Eq.~\eqref{eq:experiment}, are not all independent. They obey the Onsager relations, which can be obtained in the general framework of non-equilibrium statistical mechanics \cite{Onsager31,Onsager31a,vanVliet08}. When allowing for the presence of an external magnetic field {\bf B}, these relations read $\hat{\sigma}({\bf B})=\hat{\sigma}^T(-{\bf B})$, $\hat{\mathcal{L}}({\bf B})=\hat{\mathcal{L}}^T(-{\bf B})$, and $\hat{\mathcal{N}}({\bf B})=-T\hat{\mathcal{M}}^T(-{\bf B})$. As a direct consequence, one can obtain
\begin{align}
\hat{\rho}({\bf B})&=\hat{\rho}^T(-{\bf B}) \label{eq:Onsager_rho}{\color{red},}
\\
\hat{\kappa}({\bf B})&=\hat{\kappa}^T(-{\bf B}),\label{eq:Onsager_kappa}\\
\hat{\pi}({\bf B})&=T\hat{\alpha}^T(-{\bf B}).\label{eq:Onsager_alpha}
\end{align}
In particular, in the absence of a magnetic field, $\hat{\sigma}^T=\hat{\sigma}$, $\hat{\rho}^T=\hat{\rho}$, $\hat{\mathcal{L}}^T=\hat{\mathcal{L}}$ and $\hat{\kappa}^T=\hat{\kappa}$ are symmetric.

\section{Linearized Boltzmann equation}
\label{sec:LinBoltz}

We consider an electron system in two or three spatial dimensions with a quadratic dispersion $\epsilon_{\bf p} = p^2/(2m)$. In the presence of a magnetic field, and in linear response to an electric field or a temperature gradient, the non-equilibrium steady state is governed by the linearized Boltzmann equation, which we present in the form \cite{Ziman01}
\begin{align}
&\left( - e {\bf E} - \xi_{\bf p} \frac{\nabla T}{T} \right) \cdot {\boldsymbol v}_{\bf p} \frac{\partial n_F(\xi_{\bf p})}{\partial \xi_{\bf p}}-e ({\boldsymbol v}_{\bf p} \times {\bf B}) \cdot \nabla_{\bf p} \delta f_{\bf p}\nonumber\\
&= I_{ei}\{f\} + I_{ee}\{f\} .
\label{LinearizedBoltzmannEq_Magnetic_0}
\end{align}
Here, the distribution function has been expanded as $f({\bf r}, {\bf p}) \approx n_{F}(\xi_{{\bf p}}) + \delta f_{{\bf p}}$, where $n_{F}(\xi_{{\bf p}}) = [\exp(\beta\xi_{{\bf p}}) + 1]^{-1}$ is the Fermi-Dirac distribution with $\beta = (k_BT)^{-1}$, and $\xi_{{\bf p}} = \epsilon_{{\bf p}} - \mu$, where $\mu$ is the chemical potential. The velocity ${\boldsymbol v}_{\bf p}$ is related to the momentum ${\bf p}$ as ${\boldsymbol v}_{\bf p}={\bf p}/m$. In the two-dimensional (2$d$) case, only the magnetic field component perpendicular to the plane is effective, while ${\bf E}$ and $\nabla T$ lie in the plane. When writing the Boltzmann equation in the form of Eq.~\eqref{LinearizedBoltzmannEq_Magnetic_0}, we assumed that spin-related effects are not important in the parameter regime under consideration. This requires, in particular, that the Zeeman splitting is much smaller than the Fermi energy.

We describe the electron-impurity and electron-electron collision integrals in the relaxation-time approximation (RTA)
\begin{align}
I_{ei}\{f\}
&= - \frac{\delta f_{\bf p}}{\tau_{ei,{\bf p}}},\\
I_{ee}\{f\}
&= - \frac{f({\bf r},{\bf p}) - n_F^{({\rm cm})}(\bf p)}{\tau_{ee}}.
\label{Coll_ee}
\end{align}
There is an important difference between these two collision integrals. Impurities cause a relaxation of the electronic system towards equilibrium in the laboratory frame characterized by the distribution function $n_{F}(\xi_{\bf p})$. Electron-electron collisions conserve the total momentum of the colliding particles. Therefore, the relaxation in this case is towards equilibrium in the center of mass frame, characterized by the ``drifting distribution function" $n_F^{({\rm cm})}(\bf p)$. In contrast to the electron-impurity scattering time $\tau_{ei,{\bf p}}$, the electron-electron scattering time $\tau_{ee}$ is momentum-independent in the RTA. This ensures consistency with the conservation of momentum during electron-electron collisions. For our purposes, we assume that $\tau_{ei,{\bf p}}$ depends on $|{\bf p}|$ only. {\color{red}}

Linearizing the drifting distribution function $n_F^{({\rm cm})}({\bf p}) \approx (1 - {\boldsymbol v}_{\rm cm} \cdot {\bf p} \partial_{\xi_{\bf p}}) n_F(\xi_{\bf p})$ in $I_{ee}\{f\}$, where $\boldsymbol{v}_{\rm cm}$ is the center of mass velocity, we recast Eq.~\eqref{LinearizedBoltzmannEq_Magnetic_0} in the form
\begin{align}
&\left( - e \tilde{{\bf E}} - \xi_{\bf p} \frac{\nabla T}{T} \right) \cdot {\boldsymbol v}_{\bf p} \frac{\partial n_F(\xi_{\bf p})}{\partial \xi_{\bf p}}\nonumber\\
&= - \frac{\delta f_{\bf p}}{\tilde{\tau}_{\bf p}} + e ({\boldsymbol v}_{\bf p} \times {\bf B}) \cdot \nabla_{\bf p} \delta f_{\bf p},
\label{LinearizedBoltzmannEq_Magnetic_1}
\end{align}
with the effective electric field
\begin{align}
\tilde{{\bf E}} = {\bf E} - \frac{m{\boldsymbol v}_{\rm cm}}{e\tau_{ee}},
\label{EffectiveElectricField}
\end{align}
and the total scattering rate
\begin{align}
\frac{1}{\tilde{\tau}_{\bf p}}
= \frac{1}{\tau_{ei,{\bf p}}} + \frac{1}{\tau_{ee}}.
\label{TotalScatteringRate}
\end{align}
The inelastic scattering rate $1/\tau_{ee}$ contributes a temperature dependence to $1/\tilde{\tau}_{\bf p}$ that is not present when studying  elastic electron-impurity scattering alone. As far as the effective electric field $\tilde{\bf E}$ is concerned, the distinction between ${\bf E}$ and $\tilde{\bf E}$ becomes crucial whenever electrons acquire a finite center of mass velocity as a result of the applied electric field or temperature gradient. We see that with these definitions for the effective field $\tilde{\bf E}$ and total scattering rate $1/{\tilde{\tau}_{\bf p}}$, the linearized Boltzmann equation in the form \eqref{LinearizedBoltzmannEq_Magnetic_1} is formally equivalent to the equation governing linear response in electronic systems with only elastic scattering in the RTA. The solution of this equation is well known \cite{Beer63}. The deviation from the equilibrium distribution function $\delta f_{\bf p}$ takes the form
\begin{align}
&\delta f_{\bf p}= \frac{\tilde{\tau}_{\bf p}}{1 + (\omega_c \tilde{\tau}_{\bf p})^2} \frac{\partial n_F(\xi_{\bf p})}{\partial \xi_{\bf p}}\boldsymbol{v}_{\bf p}\cdot \label{TrialSolution_Magnetic} \\
&\left[ \big\{ 1 + \omega_c \tilde{\tau}_{\bf p} (\hat{n}_{\bf B} \times) + (\omega_c \tilde{\tau}_{\bf p})^2 \hat{n}_{\bf B} (\hat{n}_{\bf B} \cdot ) \big\} \left( e \tilde{{\bf E}} + \xi_{\bf p} \frac{\nabla T}{T} \right) \right] \nonumber,
\
\end{align}
where we defined the cyclotron frequency $\omega_c = eB/m$, and the unit vector $\hat{n}_{\bf B} = {\bf B} / |{\bf B}|$. The product of $\omega_c$ and a typical scattering time, in our case $\tilde{\tau}_{\bf p}$, frequently appears in studies of electronic transport under the influence of a magnetic field. This product describes the competition between periodic cyclotron motion and delocalizing scattering processes. The Landau level quantization is not accounted for in the Boltzmann Equation \eqref{LinearizedBoltzmannEq_Magnetic_1}. We will therefore assume that the condition $\omega_c\tilde{\tau}_{\bold{p}}<1$ holds. Under this condition, the broadening of the Landau levels caused by the scattering of various kinds is larger than the spacing between consecutive Landau levels. An additional smoothening of the Landau levels results from the motion along the magnetic field direction, if the system is three dimensional.

It is worth noting that obtaining the solution provided by Eq.~\eqref{TrialSolution_Magnetic} does not complete the problem of finding the transport coefficients. The reason is that the effective electric field $\tilde{\bf E}$ contains the center of mass velocity $\boldsymbol{v}_{\rm cm}$, which itself depends on the non-equilibrium part of the distribution function $\delta f_{\bf p}$. This dependence will be accounted for in the next section when calculating the transport coefficients self-consistently.

\section{Transport coefficients}
\label{sec:TransCoeff}

This section is concerned with the transport coefficients characterizing the electric and thermal currents flowing in response to an electric field or temperature gradient in the presence of a magnetic field ${\bf B}$ with arbitrary orientation. To find these coefficients, we make use of Eq.~\eqref{TrialSolution_Magnetic} for the non-equilibrium distribution function $\delta f_{\bf p}$ to calculate the electric and thermal currents
\begin{align}
{\bf J}_E &= - s e \int_{\bf p} \boldsymbol{v}_{\bf p} \delta f_{\bf p},\label{eq:JEbasic}\\
{\bf J}_T
 &= s \int_{\bf p} \xi_{\bf p} \boldsymbol{v}_{\bf p} \delta f_{\bf p},\label{eq:JTbasic}
\end{align}
with the particle density $\mathcal{N} = s\int_{\bf p} n_F(\xi_{\bf p})$ and the spin degeneracy $s=2$. Here, and in the following, we use the notation $\int_{\bf p}=\int d^dp/(2\pi)^d$. The right-hand side of Eq.~\eqref{TrialSolution_Magnetic} still depends on $\delta f_{\bf p}$ implicitly through the drift velocity $\boldsymbol{v}_{\rm cm}$ contained in $\tilde{\bf E}$. Indeed, the drift velocity is given as $\boldsymbol{v}_{\rm cm} = s \int_{\bf p} {\bf p}\delta f_{\bf p} / (\mathcal{N}m)$. This does not pose a problem, however, since we can eliminate $\boldsymbol{v}_{\rm cm}$ in favor of the electric current ${\bf J}_E = - \mathcal{N} e \boldsymbol{v}_{\rm cm}$. We can therefore find the currents ${\bf J}_E$ and ${\bf J}_T$ as a function of ${\bf E}$ and $\nabla T$. This leads directly to the conductivity tensors for electric, thermal and thermoelectric transport, Eq.~\eqref{eq:linear_response}. However, here we will choose a different representation that allows for a more straightforward comparison with experimental measurements, and write ${\bf E}$ and ${\bf J}_T$ as functions of ${\bf J}_E$ and $\nabla T$ as in Eq.~\eqref{eq:experiment}. Technical details of the calculation are relegated to Appendix \ref{app:derivation}. The result can be conveniently formulated by introducing the following notation
\begin{align}
\hat{\rho}&=\rho_\perp+R_H({\bf B}\times)+(\rho_\parallel-\rho_\perp) \hat{n}_{\bf B}(\hat{n}_{\bf B}\cdot),\label{eq:rhoparameter}\\
\hat{\alpha}&=S_\perp+\eta({\bf B}\times)+(S_\parallel-S_\perp)\hat{n}_{\bf B}(\hat{n}_{\bf B}\cdot),\\
\hat{\kappa}&=\kappa_\perp-\kappa\mathcal{L}({\bf B}\times)+(\kappa_\parallel-\kappa_\perp)\hat{n}_{\bf B}(\hat{n}_{\bf B}\cdot).
\end{align}
Furthermore, the Peltier coefficient tensor $\hat{\pi}$ can be eliminated in favor of $\hat{\alpha}$ by way of the relation $\hat{\pi}=T\hat{\alpha}$. The resulting equations take the form
\begin{align}
{\bf E}
&= \rho_\perp {\bf J}_E + S_\perp \nabla T + {\bf B} \times (R_H {\bf J}_E + \eta \nabla T)\nonumber\\
&+(\rho_{\parallel}-\rho_\perp)\hat{n}_{\bf B}(\hat{n}_{\bf B}\cdot {\bf J}_E)+(S_{\parallel}-S_\perp)\hat{n}_{\bf B}(\hat{n}_{\bf B}\cdot\nabla T),\label{eq:Efinal}\\
{\bf J}_T
&=TS_\perp {\bf J}_E - \kappa_\perp \nabla  T + {\bf B} \times (T\eta {\bf J}_E + \kappa_\perp \mathcal{L} \nabla T)\nonumber\\
&+T(S_{\parallel}-S_\perp))\hat{n}_{\bf B} (\hat{n}_{\bf B}\cdot {\bf J}_E)-(\kappa_\parallel-\kappa_\perp) \hat{n}_{\bf B} (\hat{n}_{\bf B}\cdot \nabla T).\label{eq:JTfinal}
\end{align}
We see that linear response transport for our isotropic model is determined by nine independent transport coefficients.

Electric transport in the absence of a temperature gradient is characterized by three coefficients, the resistivity in a perpendicular magnetic field $\rho_\perp$, the resistivity in a parallel magnetic field $\rho_\parallel$, and the Hall coefficient $R_H$,
\begin{align}
\rho_\perp&=\frac{m}{\mathcal{N}e^2}\left(\frac{Y_{00}}{Y_{00}^2+Y_{01}^2}-\frac{1}{\tau_{ee}}\right),\label{eq:rhoperp}\\
R_H&=-\frac{m}{\mathcal{N}e^2}\frac{1}{B}\frac{Y_{01}}{Y_{00}^2+Y_{01}^2},\label{eq:RH}\\
\rho_\parallel
&=\frac{m}{\mathcal{N}e^2}\left(\frac{1}{\left\langle\!\left\langle\tilde{\tau}_{\bf p}\right\rangle\!\right\rangle}-\frac{1}{\tau_{ee}}\right).\label{eq:rhoparallel}
\end{align}
In order to formulate the results in a compact form, we introduced the following matrix
\begin{align}
Y_{mn} = \bigg\langle\!\!\!\bigg\langle \frac{\xi_{\bf p}^m (\omega_c \tilde{\tau}_{\bf p})^n \tilde{\tau}_{\bf p}}{1 + (\omega_c \tilde{\tau}_{\bf p})^2} \bigg\rangle\!\!\!\bigg\rangle.
\label{Ymn}
\end{align}
Here, for any physical quantity $X_{\bf p}$, the average $\left\langle\!\left\langle \dots \right\rangle\!\right\rangle$ is defined as
\begin{align}
\langle\!\langle X_{\bf p} \rangle\!\rangle
= - \frac{2s}{d\mathcal{N}} \int_{\bf p} X_{\bf p} (\xi_{\bf p} + \mu) \frac{\partial n_F(\xi_{\bf p})}{\partial \xi_{\bf p}}.
\label{FluctuationAverage}
\end{align}
The weight $(\xi_{\bf p} + \mu)$ appearing in the definition of the average may also be expressed in terms of the square of the velocity, $\boldsymbol{v}_{\bf p}^2=2(\xi_{\bf p}+\mu)/m$.

The inelastic scattering rate $1/\tau_{ee}$ enters the Eqs.~\eqref{eq:rhoperp}-\eqref{eq:rhoparallel} in two distinct ways. First, it enters through the total scattering rate $1/\tilde{\tau}_{\bf p}$, implicitly contained in the function $Y_{mn}$. Secondly, the expressions for $\rho_\perp$ and $\rho_\parallel$ contain $1/\tau_{ee}$ explicitly. This dependence on $1/\tau_{ee}$ has its origin in the $\boldsymbol{v}_{\rm cm}$ dependence of the drifting distribution function, a dependence that arises due to the conservation of the total momentum during electron-electron collisions. It is worth mentioning that all coefficients $\rho_\perp$, $R_H$, and $\rho_\parallel$ are even in ${\bf B}$ in agreement with the Onsager relation Eq.~\eqref{eq:Onsager_rho}.

Thermoelectric transport depends on the coefficients
\begin{align}
S_\perp&=-\frac{1}{eT}\frac{Y_{00}Y_{10}+Y_{01}Y_{11}}{Y_{00}^2+Y_{01}^2},\label{eq:Sperpgeneral}\\
\eta&=-\frac{1}{eT}\frac{1}{B}\frac{Y_{00}Y_{11}-Y_{01}Y_{10}}{Y_{00}^2+Y_{01}^2},\label{eq:eta}\\
S_\parallel&=
-\frac{1}{eT}\frac{\left\langle\!\left\langle \xi_{\bf p} \tilde{\tau}_{\bf p}\right\rangle\!\right\rangle}{\left\langle\!\left\langle \tilde{\tau}_{\bf p}\right\rangle\!\right\rangle}{\color{red}.}\label{eq:Sparallelgeneral}
\end{align}
Here, $S_\perp$ ($S_\parallel$) is the Seebeck coefficient, or thermoelectric power, in perpendicular (parallel) magnetic field; $\eta$ is the Nernst coefficient. Note that $S_{\parallel}$ does not depend on the magnetic field, $S_\parallel=S_\perp(B=0)$.

In view of the Onsager relations, in combination with the relation $\hat{\pi}= T\hat{\alpha}$, we expect $\hat{\alpha}({\bf B})=\hat{\alpha}^T(-{\bf B})$, and therefore the coefficients $S_\perp$, $S_\parallel$ and $\eta$ must be even in ${\bf B}$. This property can indeed be checked from the explicit relations.

The following three coefficients determine thermal transport
\begin{align}
\kappa_\perp=&\frac{\mathcal{N}}{mT}\left(Y_{20}-\frac{Y_{00}(Y_{10}^2-Y_{11}^2)+2Y_{01}Y_{10}Y_{11}}{Y_{00}^2+Y_{01}^2}\right),\label{eq:kappa_perp}\\
\kappa_\perp\mathcal{L}=&-\frac{\mathcal{N}}{mT}\frac{1}{B}\left(Y_{21}-\frac{Y_{01}(Y_{11}^2-Y_{10}^2)+2Y_{00}Y_{10}Y_{11}}{Y_{00}^2+Y_{01}^2}\right),\label{eq:L}\\
\kappa_\parallel
=&\frac{\mathcal{N}}{mT}\left(\left\langle\!\left\langle \xi_{\bf p}^2\tilde{\tau}_{\bf p}\right\rangle\!\right\rangle-\frac{\left\langle\!\left\langle \xi_{\bf p} \tilde{\tau}_{\bf p}\right\rangle\!\right\rangle^2}{\left\langle\!\left\langle \tilde{\tau}_{\bf p}\right\rangle\!\right\rangle}\right)
.\label{eq:kappa_parallel}
\end{align}
In these equations, $\kappa_\perp$ and $\kappa_\parallel$ are the thermal conductivities in perpendicular and parallel magnetic field, respectively; $\mathcal{L}$ is the thermal Hall (or Righi-Leduc) coefficient. All coefficients are even in ${\bf B}$, in accordance with Eq.~\eqref{eq:Onsager_kappa}. Just as $\rho_\parallel$ and $S_\parallel$, $\kappa_\parallel$ does not depend on ${\bf B}$, $\kappa_\parallel=\kappa_\perp(B=0)$.

In theoretical studies, it is often easier to find the components of the generalized conductivity matrix connecting currents and external perturbations in Eq.~\eqref{eq:linear_response} than the components of the matrix of Eq.~\eqref{eq:experiment}, which is more directly related to experimental observations. Let us therefore mention here, for the example of the resistivity and conductivity tensors, the relation between the coefficients used to parameterize the matrix $\hat{\rho}$ [compare Eq.~\eqref{eq:rhoparameter}] and an analogous parameterization of the matrix $\hat{\sigma}$. Defining the coefficients $\sigma_\perp$, $\sigma_\parallel$ and $\alpha_H$, through the following equation (for $\nabla T=0$)
\begin{align}
{\bf J}_E&=\sigma_\perp {\bf E}-\alpha_H {\bf B}\times {\bf E}+(\sigma_\parallel-\sigma_\perp)\hat{n}_{\bf B}(\hat{n}_{\bf B}\cdot {\bf E}),
\end{align}
one finds the following relation between the components
\begin{align}
\sigma_\perp&=\frac{\rho_\perp}{\rho_\perp^2+\rho_H^2},\quad \sigma_\parallel=\frac{1}{\rho_\parallel},
\quad \alpha_H=\frac{1}{B}\frac{\rho_H}{\rho_\perp^2+\rho_H^2},
\end{align}
where $\rho_H=R_HB$ is the Hall resistivity.

The linear response equations Eqs.~\eqref{eq:Efinal} and \eqref{eq:JTfinal} in combination with the general expressions for the coefficients Eqs.~\eqref{eq:rhoperp}-\eqref{eq:rhoparallel} and \eqref{eq:Sperpgeneral}-\eqref{eq:kappa_parallel} are the main results of this paper. They characterize electric, thermal and thermoelectric transport accounting for electron-electron scattering and momentum-dependent electron-impurity scattering, for an arbitrary orientation of the magnetic field. In two spatial dimensions, only the magnetic field component perpendicular to the plane is effective. In this case, the coefficients $\rho_\parallel$, $S_\parallel$, and $\kappa_\parallel$ are not required for the characterization of transport, and the second lines of both Eqs.~\eqref{eq:Efinal} and \eqref{eq:JTfinal} should be discarded. Within the framework of the Boltzmann equation, these stated results are exact. Below, we will discuss the implications for different parameter regimes.

The main purpose of this manuscript is to discuss how elastic and inelastic scattering times of similar magnitude influence different transport coefficients. In principle, the formulas derived on the basis of the Boltzmann equation below are also applicable when either the elastic scattering time is much shorter than the inelastic scattering time, or in the opposite limit, which corresponds to the hydrodynamic regime. However, an important aspect relevant for the comparison with hydrodynamics is that the role of elastic scattering can be quite different for imperfections of different type. Here, we implicitly assume that the size of the impurities is smaller than both elastic and the inelastic mean free paths. This contrasts a typical hydrodynamic approach in which variations of the potential are assumed to be smooth, as for example in Ref.~\cite{Andreev11}, or scattering on a boundary is considered as, e.g., in Ref.~\cite{Levinson77}. In Ref.~\cite{Andreev11}, charge and heat transport in the presence of large-scale inhomogeneities was studied. We, in turn, consider systems in which small-size impurities distributed homogeneously in the bulk of the liquid are weak and dense.

\subsection{Dependence on the magnetic field direction}

Equations~\eqref{eq:Efinal} and \eqref{eq:JTfinal} are valid for an arbitrary direction of the magnetic field. Choosing a setup with perpendicular and parallel magnetic fields simplifies the equations and highlights the physical significance of the coefficients. We choose the direction of the electric current ${\bf J}_E$ (the temperature gradient $\nabla T$) as a reference for the direction of the magnetic field when $\nabla T=0$ (when ${\bf J}_E=0$). After discussing these two limiting cases, we consider the general case of a tilted magnetic field.

\subsubsection{Perpendicular and parallel magnetic fields}

\paragraph{${\bf B}\perp {\bf J}_E$, $\nabla T=0$:} Here, both the electric field and the thermal current are confined to the plane spanned by ${\bf J}_E$ and ${\bf B}\times{\bf J}_{E}$,\begin{align}
{\bf E}&=\rho_\perp{\bf J}_E+R_H{\bf B}\times {\bf J}_E,\\
{\bf J}_T&=TS_\perp {\bf J}_E+T\eta {\bf B}\times {\bf J}_E.
\end{align}
Neither of these quantities has a component in the direction of the magnetic field.

\paragraph{${\bf B}\perp \nabla T$, ${\bf J}_E=0$:} In this case, the electric field and thermal current both lie in the plane spanned by $\nabla T$ and ${\bf B}\times \nabla T$,
\begin{align}
{\bf E}&=S_\perp\nabla T+\eta {\bf B}\times \nabla T,\\
{\bf J}_T&=-\kappa_\perp \nabla T+\kappa_\perp \mathcal{L}{\bf B}\times \nabla T.
\end{align}

\paragraph{${\bf B}\parallel {\bf J}_E$, $\nabla T=0$:}
The magnetic field, the electric current, the thermal current and the electric field are all parallel to each other, ${\bf E}=\rho_\parallel{\bf J}_E$, ${\bf J}_T=TS_\parallel {\bf J}_E$.

\paragraph{${\bf B}\parallel \nabla T$, ${\bf J}_E=0$:}
Here, the magnetic field, the electric field, the temperature gradient and the thermal current are parallel, ${\bf E}=S_\parallel \nabla T$, ${\bf J}_T=-\kappa_\parallel \nabla T$.

\subsubsection{Tilted magnetic field}

\begin{figure}[t]
\centering
\includegraphics[width=0.31\textwidth]
{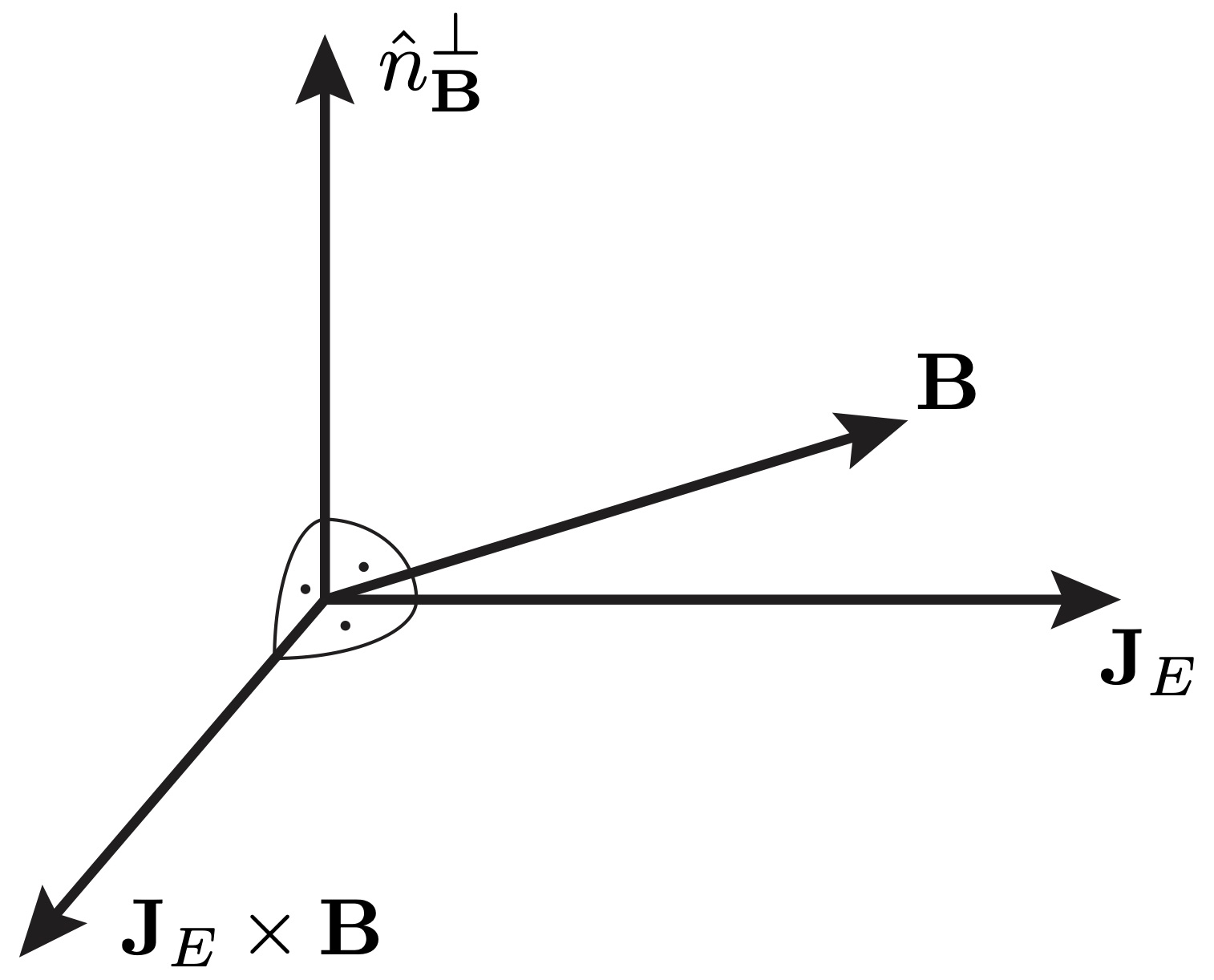} \\
\caption{This figure shows the different vectors that are important for the discussion of electric transport for the general tilted magnetic field case in three dimensions, when ${\bf B}$ is neither perpendicular nor parallel to ${\bf J}_E$. In this case, an electric field component parallel to $\hat{n}_{\bf B}^\perp\propto {\bf J}_E\times({\bf B}\times {\bf J}_E)$ arises, i.e., the electric field has a component pointing out of the plane spanned by ${\bf B}$ and ${\bf J}_E\times{\bf B}$, unlike for the perpendicular magnetic field case. This component along $\hat{n}_{\bf B}^\perp$ is non-vanishing only for $\rho_\parallel\ne \rho_\perp$, as can be seen from Eq.~\eqref{eq:Edecomp}.  }
\label{Figure1}
\end{figure}

An interesting observation can be made for the case of an arbitrary magnetic field direction, see Fig.~\ref{Figure1}. For the purpose of illustration we highlight the case of electric transport in the absence of a temperature gradient, which is characterized by the three coefficients $\rho_\perp$, $\rho_\parallel$ and $R_H$. Alternative setups involving the other transport coefficients can be discussed in analogy. For the case under consideration, we have
\begin{align}
{\bf E}=\rho_\perp{\bf J}_E+R_H{\bf B}\times {\bf J}_E+(\rho_\parallel-\rho_\perp)\hat{n}_{\bf B}(\hat{n}_{\bf B}\cdot {\bf J}_E){\color{red}.}
\label{eq:forE}
\end{align}
We see that for an arbitrary direction of the magnetic field, $\hat{n}_{\bf B}$ can have a component in the direction of the electric current. This observation motivates the following decomposition
\begin{align}
\hat{n}_{\bf B}&=\hat{n}_{\bf B}^\parallel+\hat{n}_{\bf B}^\perp,\\
\hat{n}_{\bf B}^\parallel&=\hat{n}_{{\bf J}_E}(\hat{n}_{{\bf J}_E}\cdot \hat{n}_{\bf B}),\\
\hat{n}_{\bf B}^\perp&=\hat{n}_{{\bf J}_E}\times (\hat{n}_{\bf B}\times \hat{n}_{{\bf J}_E}),
\end{align}
where $\hat{n}_{\bf B}^\parallel$ is parallel to the electric current, and $\hat{n}_{\bf B}^\perp$ is perpendicular. This allows us to bring Eq.~\eqref{eq:forE} into the form
\begin{align}
{\bf E}&=\left[\rho_\perp+(\hat{n}_{\bf B}^\parallel)^2(\rho_\parallel-\rho_\perp)\right]{\bf J}_E+R_H{\bf B}\times {\bf J}_E\nonumber\\
&+(\rho_\parallel-\rho_\perp)(\hat{n}_{\bf B}^\parallel\cdot {\bf J}_E)\hat{n}_{\bf B}^\perp{\color{red},}
\label{eq:Edecomp}
\end{align}
with three mutually orthogonal vectors ${\bf J}_E$, ${\bf B}\times {\bf J}_E$, and $\hat{n}_{\bf B}^\perp$. This is illustrated in Fig.~\ref{Figure1}. The last term is relevant only if ${\bf J}_E$ and ${\bf B}$ are neither parallel nor perpendicular, i.e. for general tilted magnetic fields. Then, the electric field develops a component perpendicular to both ${\bf J}_E$ and ${\bf B}\times{\bf J}_E$, which is proportional to $\rho_\parallel-\rho_\perp$.

\subsection{Constant elastic scattering rate}
\label{constant_tau}

If we eliminate the momentum dependence of $\tau_{ei}$, $\tilde{\tau}_{\bf p}\rightarrow \tilde{\tau}$, then we find
\begin{align}
\rho_\perp&=\rho_\parallel=\rho_0,\quad R_H=R_{H0}{\color{red},}
\\
S_\perp&=S_\parallel=S_0,\quad \eta=0{\color{red},}
\\
\kappa_\perp&=\kappa_0,\quad \kappa_\parallel=\kappa_0\left(1+(\omega_c\tilde{\tau})^2\right),\quad
\mathcal{L}=\mathcal{L}_0{\color{red},}
\end{align}
where
\begin{align}
\rho_0&=\frac{m}{\mathcal{N}e^2\tau_{ei}},
~~R_{H0}=-\frac{1}{\mathcal{N}e},
~~S_0=-\frac{1}{eT}\left\langle\!\left\langle \xi_{\bf p}\right\rangle\!\right\rangle\label{eq:0values}{\color{red},}
\\
\kappa_0&=\frac{\mathcal{N}}{mT}\frac{\tilde{\tau}}{1+(\omega_c\tilde{\tau})^2}\left(\left\langle\!\left\langle \xi_{\bf p}^2\right\rangle\!\right\rangle-\left\langle\!\left\langle \xi_{\bf p}\right\rangle\!\right\rangle^2\right),\; \mathcal{L}_0=-\frac{e\tilde{\tau}}{m}.\nonumber
\end{align}
At low temperatures $T\ll \epsilon_F$, the two moments of $\xi_{\bf p}$ entering the expressions for $\kappa_0$ and $S_0$ are $\left\langle\!\left\langle \xi_{\bf p}^2\right\rangle\!\right\rangle=\pi^2 T^2/3$ and $\left\langle \!\left\langle\xi_{\bf p}\right\rangle\!\right\rangle=\pi^2 T^2/2\epsilon_F$.

A few remarks are in order here. For a constant scattering time $\tau_{ei}$, the coefficients $\rho_0$, $R_{H0}$, $S_0$, and $\eta$ do not depend on $\tau_{ee}$. In this limit, the Fermi sphere is shifted as a whole under the influence of the electric field. As a consequence, inelastic scattering becomes ineffective for the conductivity tensor $\hat{\sigma}$, cross effect tensor $\hat{\mathcal{N}}$, and the Onsager related $\hat{\mathcal{M}}$. This argument does not hold for the components of the thermal conductivity tensor due to the additional factor $\xi_{\bf p}$ associated with the temperature gradient in Eq.~\eqref{TrialSolution_Magnetic}. Furthermore, for the thermal conductivity, a difference between parallel $\parallel$ and perpendicular $\perp$ components remains in the limit of constant $\tau_{ei}$, i.e., $\kappa_\parallel \ne \kappa_\perp$. This leads to a nontrivial angular dependence as can be seen from the thermal {\it{analog}} of Eq.~\eqref{eq:Edecomp}. Finally, $\kappa_\perp$ is the only coefficient that depends on $B$ for a constant elastic scattering rate.

\section{Discussion}
\label{sec:discussion}

The temperature dependence of the transport coefficients originates from two sources. First, from the thermal smearing encoded in the averages $\left\langle\!\left\langle \dots\right\rangle\!\right\rangle$ defined in Eq.~\eqref{FluctuationAverage}, and secondly from the temperature dependence of the inelastic scattering rate. The latter is a phenomenological parameter and needs to be fixed externally. For a momentum-independent elastic scattering rate, the results for the transport coefficients simplify considerably, see Sec.~\ref{constant_tau}. As we will discuss below, it is often the momentum dependence of the elastic scattering rate that induces interesting dependences of the coefficients on temperature and magnetic field. For the sake of the discussion, we therefore single out the momentum-dependent part of the elastic scattering rate
\begin{align}
1/\tau_{ei,{\bf p}}=1/\tau_{ei}+\delta \Gamma_{\bf p}.\label{eq:fullrate}
\end{align}
The entire dependence of the transport coefficients on $\delta\Gamma_{\bf p}$ is encoded in the averages $Y_{mn}$, as can be seen from Eqs.~\eqref{eq:rhoperp}-\eqref{eq:kappa_parallel}. The momentum dependence of the elastic scattering rate enters these averages in the form of the combination $1/\tilde{\tau}_{\bf p}=1/\tilde{\tau}+\delta \Gamma_{\bf p}$. When changing the temperature, two competing trends influence $Y_{mn}$. Typically, $1/\tau_{ee}$ increases with increasing temperature. Then, the momentum dependence of $\delta \Gamma_{\bf p}$ becomes less important in comparison to the total scattering rate and so does its influence on $Y_{mn}$ and the transport coefficients. On the other hand, a larger range of momenta is probed in $Y_{mn}$ as the temperature increases due to the weighting factor $\partial n_F(\xi_{\bf p})/\partial \xi_{\bf p}$ entering the averages [compare Eq.~\eqref{FluctuationAverage}]. This effect enhances the influence of $\delta \Gamma_{\bf p}$ when the temperature grows.

Figs.~\ref{fig:rhoperp}-\ref{Figure5} illustrate the temperature and magnetic field dependence of the transport coefficients $\rho_\perp$, $S$, $\eta$, $\rho_\parallel=\rho_\perp(B=0)$, and $S_\parallel=S_\perp(B=0)$. In addition, Figs.~S1-S3 display $R_H$, $\kappa_\perp$, $\kappa_\parallel=\kappa_\perp(B=0)$, and $\mathcal{L}$ \cite{Suppl}. Knowledge of these coefficients is sufficient for the characterization of transport in a magnetic field of arbitrary direction. In all figures, we apply the same notation. Solid lines illustrate results for a momentum-dependent elastic scattering rate. Solid black, red and blue lines are computed for $\omega_c\tau_{ei}=0$, 0.4, 0.8, respectively. Dashed lines are calculated for a momentum-independent scattering rate. As we have already mentioned, the thermal conductivity $\kappa_\perp$ is the only coefficient that depends on the magnetic field even for a constant elastic scattering rate. For the purpose of our illustrations, we assume that the electron-electron scattering rate has a quadratic dependence on temperature, as in a Fermi liquid at low temperatures, with $1/\tau_{ee}=3.44\times T^2/\epsilon_F$. We also assume that disorder is weak, and set $1/(\epsilon_F\tau_{ei})=0.01$. For the illustrations of the results, we choose the parameterization
\begin{align}
\tau_{ei}\delta \Gamma_{\bf p}=w_1\xi_{\bf p}/\epsilon_F+w_2(\xi_{\bf p}/\epsilon_F)^2\label{eq:deltaGamma}
\end{align}
with $w_1=2.3$ and $w_2=1.4$, and set the dimensionality to $d=3$. For the interpretation of the results, it will be instructive to expand the expressions for the transport coefficients in powers of the momentum-dependent part of the elastic scattering rate, $\delta \Gamma_{\bf p}$. Next, we will discuss the characteristics of the transport coefficients $\rho_\perp$, $\rho_\parallel$, $S_\perp$, $S_\parallel$, and $\eta$. A discussion of the coefficients $R_H$, $\kappa_\perp$, $\kappa_\parallel$, and $\mathcal{L}$ is provided in the supplementary material \cite{Suppl}.

\subsection{Resistivity $\rho_\perp$}
\label{subsec:rhoperp}

\begin{figure}[t]
\centering
\includegraphics[width=0.45\textwidth]
{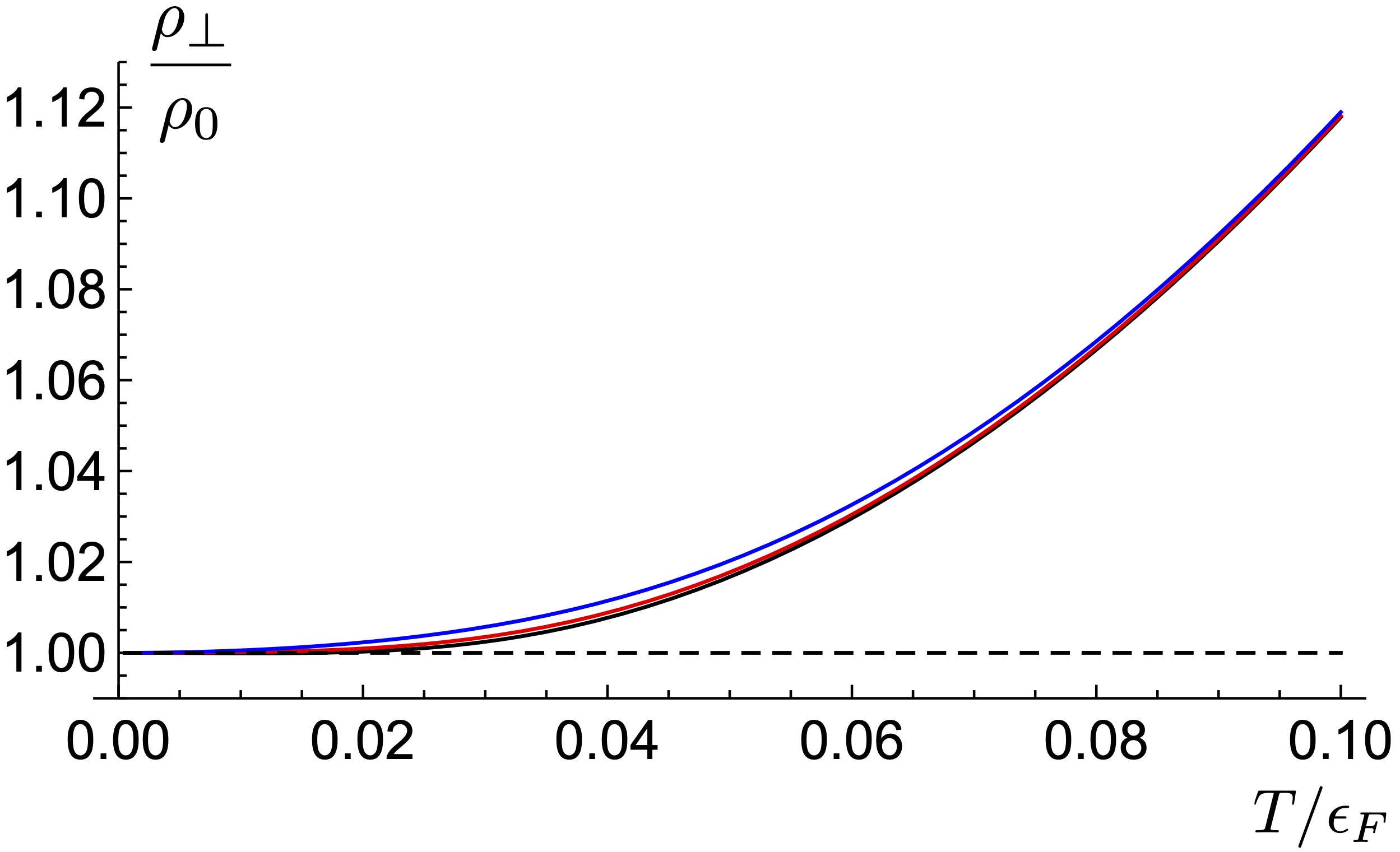} \\
\caption{
The resistances $\rho_\perp$, Eq.~\eqref{eq:rhoperp}, and $\rho_\parallel=\rho_\perp(B=0)$, Eq.~\eqref{eq:rhoparallel}, normalized to the Drude result $\rho_0$, Eq.~\eqref{eq:0values}, as a function of temperature $T$. Solid lines illustrate results for a momentum-dependent elastic scattering rate, parameterized according to Eqs.~\eqref{eq:fullrate} and \eqref{eq:deltaGamma} with $1/(\epsilon_F\tau_{ei})=0.01$, $w_1=2.3$ and $w_2=1.4$. The inelastic scattering rate is chosen as $1/\tau_{ee}=3.44\times T^2/\epsilon_F$. The dimensionality is $d=3$. Solid black, red and blue lines are computed for $B=0$, $\omega_c\tau_{ei}=0.4$, and $\omega_c\tau_{ei}=0.8$, respectively. The dashed line is calculated for a momentum-independent scattering rate, and coincides with $\rho_0$. A detailed discussion of the results is provided in Sec.~\ref{subsec:rhoperp}.
}
\label{fig:rhoperp}
\end{figure}

For a constant elastic scattering rate, the resistivity $\rho_\perp$ depends neither on  temperature nor on the magnetic field. The inelastic scattering time $\tau_{ee}$ drops out in this case and only the elastic scattering time $\tau_{ei}$ enters the expression for $\rho_\perp$,
\begin{align}
\rho_\perp\rightarrow \rho_0=\frac{n}{\mathcal{N}e^2\tau_{ei}},\quad \tau_{ei}=\mbox{const.} \label{eq:rho_0}
\end{align}

A temperature dependence arises for $\rho_\perp$ when the elastic scattering rate becomes momentum-dependent, $\delta \Gamma_{\bf p}\ne 0$. For $B=0$, this case has first been discussed by Keyes \cite{Keyes58},
\begin{align}
\rho_\perp&\rightarrow \frac{m}{\mathcal{N}e^2}\left(\frac{1}{\left\langle\!\left\langle \tilde{\tau}_{\bf p}\right\rangle\!\right\rangle}-\frac{1}{\tau_{ee}}\right),\qquad B=0,
\end{align}
with notable limits $\rho_\perp\rightarrow {m}/{\mathcal{N}e^2}\times \langle\!\langle \tau_{ei,{\bf p}}^{-1}\rangle\!\rangle$ for $\tau_{ee}\rightarrow 0$ and $\rho_\perp\rightarrow {m}/{\mathcal{N}e^2}\times \left\langle\!\left\langle \tau_{ei,{\bf p}}\right\rangle\!\right\rangle^{-1}$ for $\tau_{ee}\rightarrow \infty$.

In this paper, we focus on the low-temperature regime, $T\ll \epsilon_F$, while the relation between $1/\tau_{ei}$ and $1/\tau_{ee}$ remains arbitrary. Fig.~\ref{fig:rhoperp} shows $\rho_\perp$ for different temperatures and magnetic fields. The zeroth-order term, $\rho_0$, with respect to the momentum-dependent part of the elastic scattering rate, $\delta \Gamma_{\bf p}$ has already been discussed and is displayed in Eq.~\eqref{eq:rho_0}. The first-order term reads
\begin{align}
\delta \rho^{(1)}_\perp&=\frac{m}{\mathcal{N}e^2}\left\langle\!\left\langle \delta \Gamma_{\bf p}\right\rangle\!\right\rangle.
\end{align}
This term gives rise to a leading quadratic temperature dependence for the form of $\delta\Gamma_{\bf p}$ used here, which is clearly visible in Fig.~\ref{fig:rhoperp}. Further, we note that $\delta \rho^{(1)}_\perp$ does not depend on the magnetic field, which explains the weak magnetic field dependence observed in Fig.~\ref{fig:rhoperp}.

\subsection{Seebeck coefficient $S_\perp$}
\label{subsec:Sperp}
For a momentum-independent elastic scattering rate, the Seebeck coefficient $S_\perp$ depends neither on the magnetic field, nor on any scattering mechanism,
\begin{align}
S_\perp\rightarrow S_0=-\frac{1}{eT}\left\langle\!\left\langle \xi_{\bf p}\right\rangle\!\right\rangle, \qquad \tau_{ei}=\mbox{const}.
\end{align}
$S_{\perp}$ remains temperature-dependent in this case; for example, $S_{\perp}\propto T$ at low temperatures $T\ll \epsilon_F$. It is worth noting that $S_0$ is finite only due to particle-hole asymmetry, for which there are two sources in the model under consideration. The first one is the $\xi_{\bf p}$ dependence of $v_{\bf p}^2\propto \xi_{\bf p}+\mu$, which enters the definition of the average, Eq.~\eqref{FluctuationAverage}. The second source of particle-hole asymmetry is the $\xi_{\bf p}$ dependence of the density of states in three dimensions which becomes explicit upon changing the integration variable from ${\bf p}$ to $\xi_{\bf p}$ in Eq.~\eqref{FluctuationAverage}. For a vanishing magnetic field, but general $\tau_{ei,{\bf p}}$, one obtains
\begin{align}
S_\perp\rightarrow -\frac{1}{eT}\frac{\left\langle\!\left\langle \xi_{\bf p}\tilde{\tau}_{\bf p}\right\rangle\!\right\rangle}{\left\langle\!\left\langle \tilde{\tau}_{\bf p}\right\rangle\!\right\rangle},\qquad B=0,
\end{align}
with limiting cases $S_\perp=-\left\langle\!\left\langle \xi_{\bf p}\tau_{ei,{\bf p}}\right\rangle\!\right\rangle/(eT\left\langle\!\left\langle \tau_{ei,{\bf p}}\right\rangle\!\right\rangle)$ for $\tau_{ee}\gg \tau_{ei,{\bf p}}$ and $S_\perp=-\left\langle\!\left\langle \xi_{\bf p}\right\rangle\!\right\rangle/eT $ for $\tau_{ee}\ll \tau_{ei,{\bf p}}$. The key features of this expression have already been discussed in Ref.~\cite{Lee20}.

In order to explore the {\it combined} effect of the magnetic field and the momentum dependence of $\tau_{ei,{\bf p}}$ on the Seebeck coefficient, we expand Eq.~\eqref{eq:Sperpgeneral} up to linear order in $\delta \Gamma_{\bf p}$. The first order correction in $\delta \Gamma_{\bf p}$ reads as
\begin{align}
\delta S^{(1)}_\perp&=\frac{1}{eT}\frac{\left\langle\!\left\langle \xi_{\bf p}\delta \Gamma_{\bf p}\right\rangle\!\right\rangle-\left\langle\!\left\langle  \xi_{\bf p}\right\rangle\!\right\rangle\left\langle\!\left\langle \delta \Gamma_{\bf p}\right\rangle\!\right\rangle}{\tau_{ei}^{-1}+\tau_{ee}^{-1}}\frac{1}{1+(\omega_c\tilde{\tau})^2}.\label{eq:dS1}
\end{align}
Unlike for the resistance $\rho_\perp$, $\delta \Gamma_{\bf p}$ induces a sensitivity of $S_\perp$ to electron-electron collisions, disorder, and the magnetic field already at linear order in the expansion. For the further discussion, it is convenient to write the expression for $\delta S^{(1)}_\perp$ as the product of two factors
 \begin{align}
 \delta S^{(1)}_\perp=\delta S^{(1)}_\perp(B=0) [1+(\omega_c\tilde{\tau})^2]^{-1}.\label{eq:S1perpfactorized}
 \end{align}
The first factor, $\delta S^{(1)}_\perp(B=0)$, stands for the correction to the Seebeck coefficient in the absence of a magnetic field. The second factor encodes the entire magnetic field dependence. As discussed in Ref.~\cite{Lee20}, the correction $\delta S^{(1)}_\perp$ can give rise to a non-monotonic temperature dependence of the Seebeck coefficient $S_\perp$ for $B=0$. Let us briefly recall the argument. At low temperatures, $T\ll \epsilon_F$, and for $B=0$, the correction to the Seebeck coefficient (in $d=2,3$ dimensions) becomes
\begin{align}
\frac{\delta S_\perp^{(1)}(B=0)}{S_0}=-\frac{\frac{2}{d}w_1+\frac{16}{15}w_2\left(\frac{\pi T}{\epsilon_F}\right)^2}{1+{\tau_{ei}}/{\tau_{ee}}}.
\end{align}
The leading temperature dependence, i.e. the $w_1$ term in the low-temperature expansion, originates from the term $\left\langle\!\left\langle \xi_{\bf p}\delta \Gamma_{\bf p}\right\rangle\!\right\rangle$ in Eq.~\eqref{eq:dS1}.

\begin{figure}[tb]
\centering
\includegraphics[width=0.45\textwidth]
{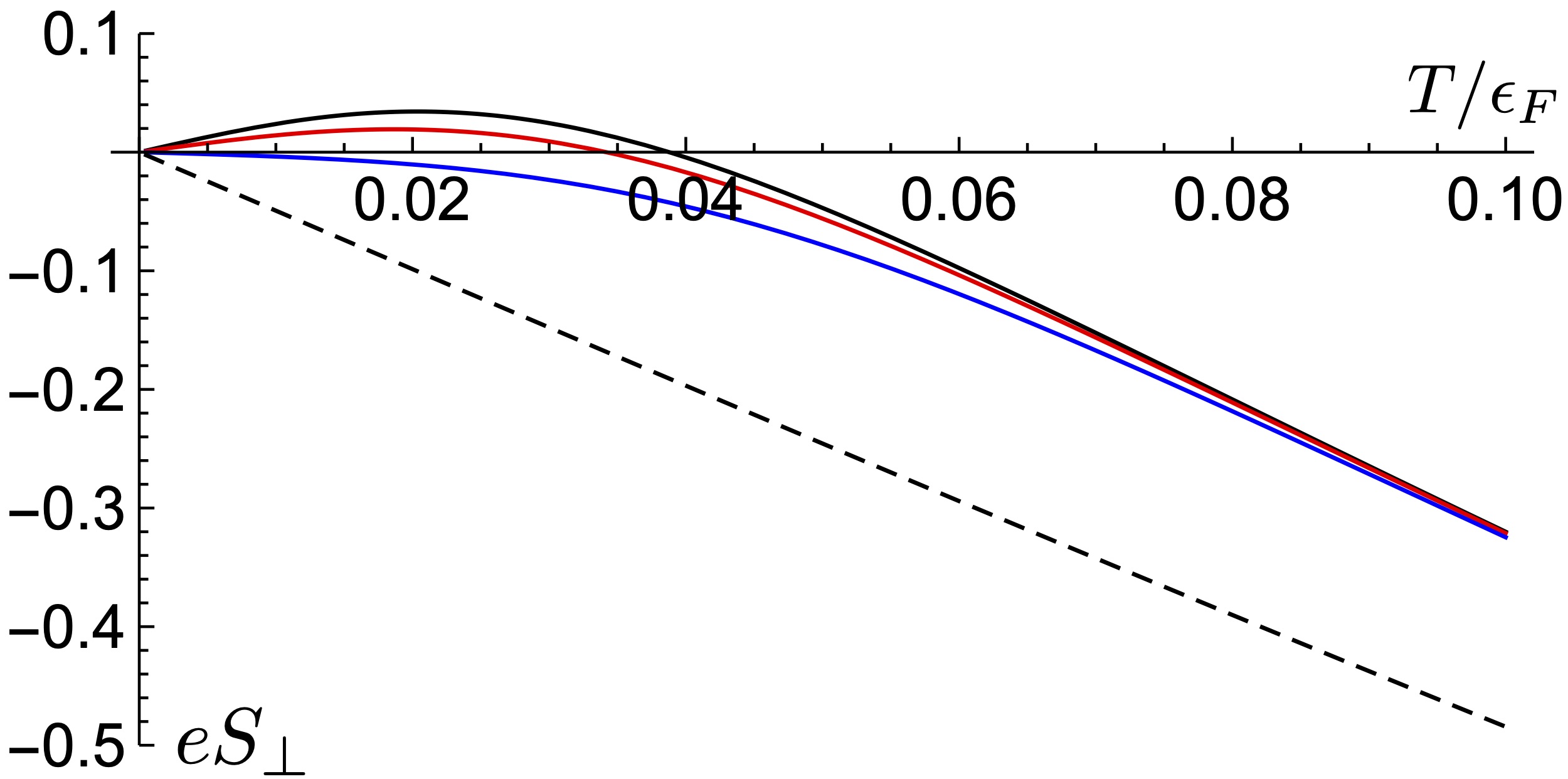} \\
\caption{
The Seebeck coefficient $S_\perp$, Eq.~\eqref{eq:Sperpgeneral}, as a function of temperature $T$, and for different magnetic fields. Parameters as listed in the caption of Fig.~\ref{fig:rhoperp}. In particular, the black solid line stands for $S_\perp(B=0)=S_\parallel$. A discussion is provided in Sec.~\ref{subsec:Sperp}.
}
\label{fig:Sperp}
\end{figure}

An important observation is that $\delta S^{(1)}_\perp$ is not necessarily smaller than $S_0$. This is because both depend on particle-hole asymmetry. For $S_0$, this dependence reveals itself through the average $\left\langle\!\left\langle \xi_{\bf p}\right\rangle\!\right\rangle$, which is finite due to a non-constant density of states and/or a non-constant velocity, as discussed above. $\delta S^{(1)}_\perp$, on the other hand, is finite due to the momentum dependence of the elastic scattering rate. In three dimensions, for example, the origin of this momentum dependence may also be the density of states, just as for $S_0$. The natural behavior in this case is $w_1>0$ and as a consequence $\delta S_\perp^{(1)}$ and $S_0$ have opposite signs at low temperatures. The expected temperature dependence of $S_\perp$ is as follows: $S_\perp$ vanishes for $T\rightarrow 0$. At low but finite temperatures, $S_\perp$ may be positive, if $\delta S^{(1)}$ dominates. For higher temperatures $\tau_{ee}$ becomes shorter, resulting in a suppression of $\delta S^{(1)}$. As a consequence, $S_\perp$ displays a maximum, and subsequently changes sign  to become negative at higher temperatures, just as the now dominant $S_0$. The temperature scale at which the maximum of $S_\perp$ occurs can be estimated by equating $\tau_{ee}$ and $\tau_{ei}$. In the presence of the magnetic field, $\delta S^{(1)}_\perp$ is suppressed by the factor $[1+(\omega_c\tilde{\tau})^2)]^{-1}$[compare Eq.~\eqref{eq:S1perpfactorized}]. The suppression becomes stronger for higher magnetic fields. As a consequence, a sign change of $S_\perp$ now requires the more stringent condition $w_1>2/d\times [1+(\omega_c\tau_{ei})^2]$. The influence of the magnetic field is most pronounced at low temperatures. Indeed, if $\tau_{ee}$ decreases with increasing temperature, which is the natural behavior, so does the product $\omega_c\tilde{\tau}$.

The temperature and magnetic field dependence of $S_\perp$ as obtained from Eq.~\eqref{eq:Sperpgeneral} is illustrated in Fig.~\ref{fig:Sperp}. We see that in accordance with our discussion (i) for $B=0$ the Seebeck coefficient displays a non-monotonic temperature dependence with a maximum at finite temperatures and a sign change, (ii) a sufficiently large magnetic field suppresses the maximum, (iii) the influence of the magnetic field decreases with increasing temperature, and (iv) $\delta S_{\perp}^{(1)}/S_0$ decreases with increasing temperature. One may, thus, conclude that measuring temperature dependence of the Seebeck coefficient at various magnetic fields provides an effective tool for determining the magnitude of the electron-impurity and electron-electron scattering rates.

\begin{figure}[tb]
\centering
\includegraphics[width=0.38\textwidth]
{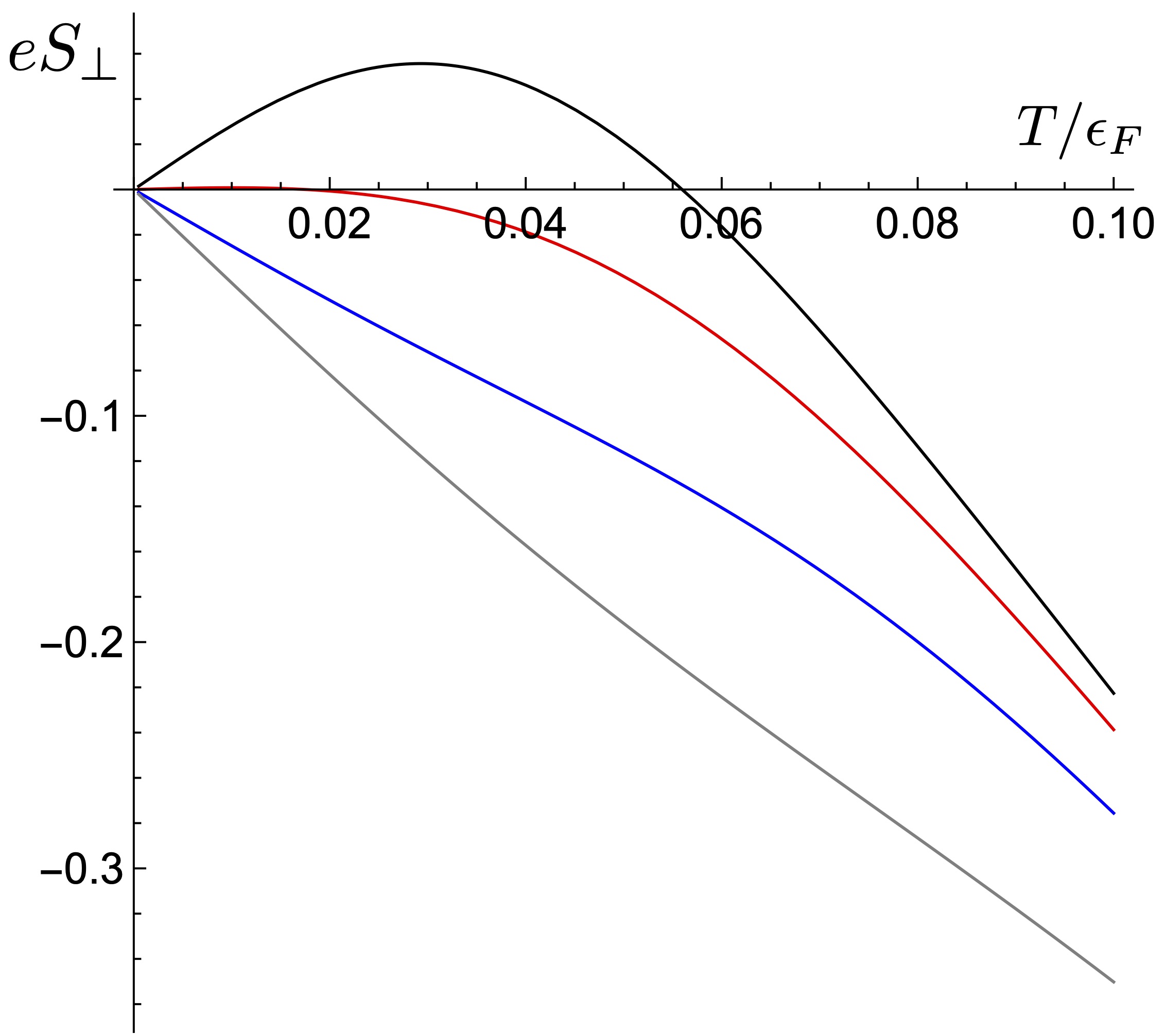} \\
\caption{The Seebeck coefficient $S_\perp$, Eq.~\eqref{eq:Sperpgeneral}, as a function of temperature $T$, and for different magnetic fields. In comparison to Fig.~\ref{fig:Sperp}, the parameters are: $1/\epsilon_F\tau_{ei}=0.05$, $w_1=2.35$, $w_2=1.4$, $1/\tau_{ee}=9.1\times T^2/\epsilon_F$. Black, red, blue and gray curves stand for $\omega_c\tau_{ei}=0$, $0.75$, $1.5$, and $3$, respectively. A discussion is provided in Sec.~\ref{subsec:Sperp}.
}
\label{fig:Sperp1}
\end{figure}

Reference ~\cite{Lakner93} reported a measurement of the Seebeck coefficient in Si:P near the $3d$ metal-insulator transition. This experiment was performed at very low temperatures $<1K$ in order to minimize the influence of phonons. Due to the closeness to the metal-insulator transition, electron-electron interactions are expected to be strong. On the metallic side of the transition, the Seebeck coefficient displays a non-monotonic temperature dependence qualitatively similar to the one discussed above. Moreover, a suppression of the maximum is observed at finite magnetic fields, eventually leading to an almost linear temperature dependence at the highest magnetic fields in the experiment. The authors of Ref.~\cite{Lakner93} interpret the observed behavior in terms of the Kondo effect  (Ref.~\cite{Maki69}) that may arise close to the metal-insulator transition due to the formation of magnetic moments. Motivated by the experimental observations, we display the Seebeck coefficient as calculated from Eq.~\eqref{eq:Sperpgeneral} once more in Fig.~\ref{fig:Sperp1}. Compared to Fig.~\ref{fig:Sperp}, both the electron-impurity and electron-electron scattering rates are increased (in relation to the Fermi energy) in Fig.~\ref{fig:Sperp1}, and higher values of $\omega_c\tau_{ei}$ are included. [Note that in the latter case the Landau level quantization may become relevant before thermal smearing smooths out quantization effects with increasing temperature.] We see that the main features of the experiment are well reproduced. Unfortunately, a direct comparison to the experiment is difficult due to the uncertainty in the relevant energy scales. We can conclude, however, that Eq.~\eqref{eq:Sperpgeneral} provides a good phenomenological description of the observed behavior.

\subsection{Nernst coefficient $\eta$}
\label{subsec:eta}

\begin{figure}[tbp]
\centering
\includegraphics[width=0.45\textwidth]
{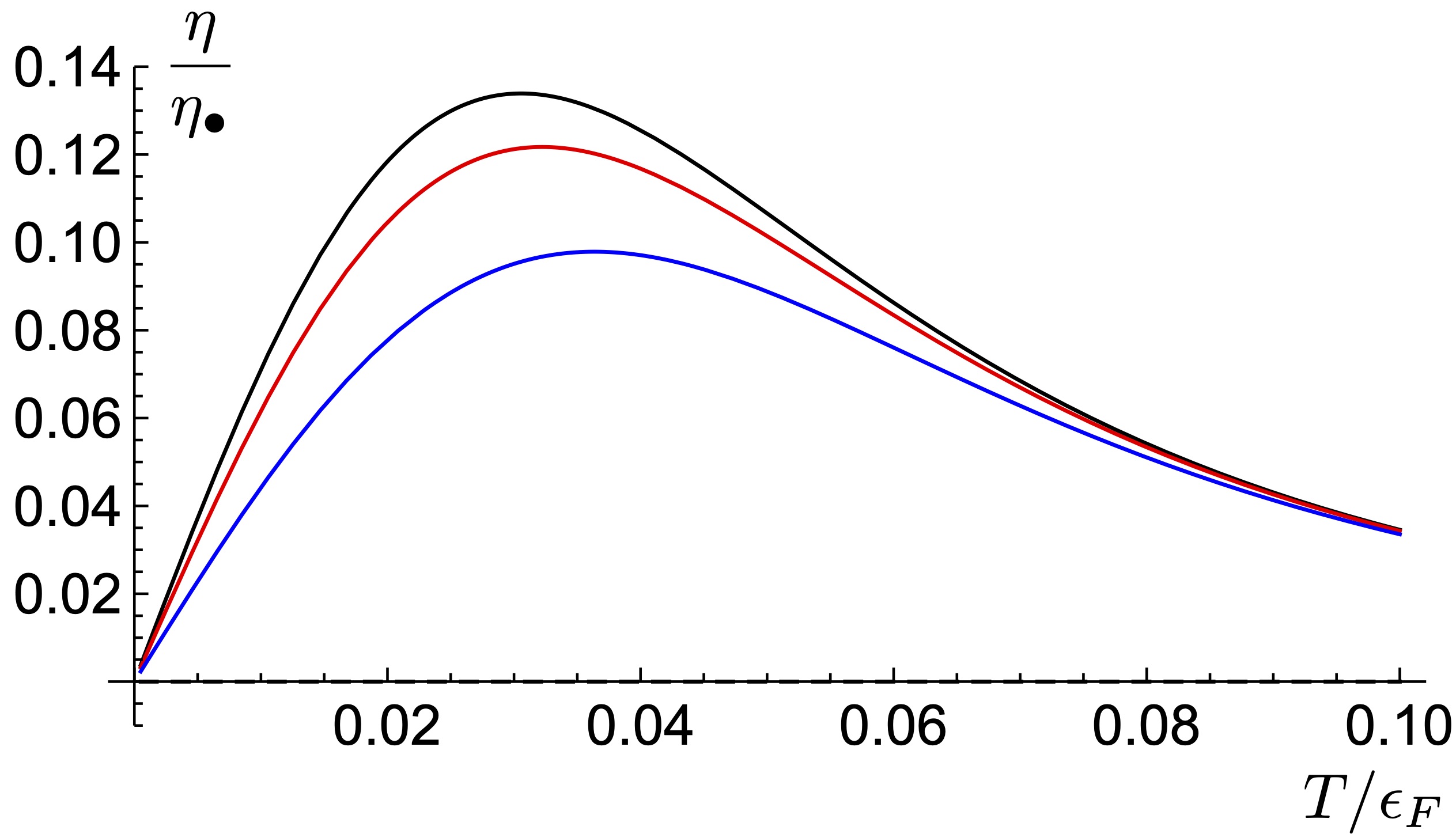} \\
\caption{The Nernst coefficient $\eta$, Eq.~\eqref{eq:eta}, normalized to $\eta_\bullet=\tau_{ei}/m$, as a function of temperature $T$, and for different magnetic fields. Parameters as listed in the caption of Fig.~\ref{fig:rhoperp}. A discussion is provided in Sec.~\ref{subsec:eta}.
}
\label{Figure5}
\end{figure}

The Nernst coefficient $\eta$ can be discussed along similar lines as $S_\perp$. An obvious difference is that the Nernst coefficient vanishes for arbitrary magnetic fields when $\tau_{ei}$ is constant. A finite Nernst coefficient is obtained, however, when the momentum dependence of $\tau_{ei,{\bf p}}$ is accounted for. At first order in $\delta \Gamma_{\bf p}$ one finds
\begin{align}
\delta \eta^{(1)}=\frac{1}{eTB}\frac{\omega_c\tilde{\tau}^2}{1+\omega_c^2\tilde{\tau}^2}\left(\left\langle\!\left\langle \xi_{\bf p}\delta \Gamma_{\bf p}\right\rangle\!\right\rangle-\left\langle\!\left\langle  \xi_{\bf p}\right\rangle\!\right\rangle\left\langle\!\left\langle \delta \Gamma_{\bf p}\right\rangle\!\right\rangle\right).
\end{align}
We have already encountered the combination of averages in round brackets in the expression for $S_{\perp}^{(1)}$ given in Eq.~\eqref{eq:dS1}. At low temperatures, the leading contribution comes from the first term, $\left\langle\!\left\langle \xi_{\bf p}\delta \Gamma_{\bf p}\right\rangle\!\right\rangle \propto w_1 T^2$, which implies that the Nernst coefficient is proportional to $T$. For $B\rightarrow 0$ the only other source of temperature dependence comes from the factor $\tilde{\tau}^2$. This factor is approximately constant at low $T$ when $\tau_{ee}\gg \tau_{ei}$, and decreases at higher temperatures, when $\tau_{ee}\ll \tau_{ei}$ and $\tilde{\tau}^2\sim \tau_{ee}^2$. Consequently, the Nernst coefficient is positive and goes through a maximum at finite $T$. A rough estimate for the temperature scale at which the maximum occurs is obtained from the condition $\tau_{ei}=\tau_{ee}$. The magnetic field dependence of $\delta \eta^{(1)}$ is governed by the factor $[1+\omega_c^2\tilde{\tau}^2]^{-1}$, which equals $[1+\omega_c^2\tau_{ei}^2]^{-1}$ for $T\rightarrow 0$, and then successively approaches $1$, when $\tilde{\tau}^2$ diminishes with increasing temperature. Therefore, the magnetic field dependence is most pronounced at low temperatures and becomes weak for $\tau_{ee}\ll \tau_{ei}$. All the described features are visible in Fig.~\ref{Figure5}, which is obtained directly from the exact result in Eq.~\eqref{eq:eta}.

\section{Conclusion}
\label{sec:conclusion}

In this paper, we studied the combined effect of electron-electron and electron-impurity scattering on charge and heat transport in metallic systems at intermediate temperatures. We employed a simple kinetic equation approach, in which both collision integrals are treated in the relaxation time approximation, and studied the linear response of the system. We found expressions for all relevant transport coefficients in the presence of a magnetic field of arbitrary direction, and analyzed the influence of the momentum dependence of the electron-impurity scattering time in detail. The results are applicable for two and three-dimensional systems.

Despite its simplicity, the model used in this paper captures a key element of the kinetics of disordered electronic systems: the competition between the relaxation of the distribution function towards equilibrium in the laboratory frame caused by electron-impurity scattering, and the relaxation towards the drifting distribution function resulting from the electron-electron interaction. This drift enters the linearized kinetic equation through the center of mass velocity $\boldsymbol{v}_{cm}$. It is straightforward to follow the effect of a finite center of mass velocity on the transport coefficients in this approach, because $\boldsymbol{v}_{cm}$ is accompanied by an explicit factor of the electron-electron scattering rate $1/\tau_{ee}$, as can be seen from Eq.~\eqref{EffectiveElectricField}. Out of the three independent tensors $\hat{\rho}$, $\hat{\alpha}$, and $\hat{\kappa}$, only $\hat{\rho}$ is affected by the finite drift velocity, while the others depend on $\tau_{ee}$ only via the total scattering rate $1/\tilde{\tau}_{\bf p}$. It should be emphasized, however, that the situation is quite different when the conductivity tensor $\hat{\sigma}$, thermal flow tensor $\hat{\mathcal{L}}$, and cross effect tensors $\hat{\mathcal{N}}$ and $\hat{\mathcal{M}}$ are used for characterizing the transport processes. These are all affected by the drift.

\begin{table}[tb]
\centering
\begin{tabular}{l l l}
\hline\hline \noalign{\smallskip}
 {\bf Resistivity} \qquad
 & $\rho_\perp$
 &
 Eq.~\eqref{eq:rhoperp}, Fig.~\ref{fig:rhoperp} \\
 \noalign{\smallskip}
\cline{2-3}\noalign{\smallskip}
& $\rho_\parallel$ \qquad
&
Eq.~\eqref{eq:rhoparallel}, Fig.~\ref{fig:rhoperp}
 \\
 \noalign{\smallskip}
\cline{1-3}\noalign{\smallskip}
{\bf Hall coefficient}\qquad & $R_H$ & Eq.~\eqref{eq:RH}, Fig.~S1\\
 \noalign{\smallskip}
\cline{1-3}\noalign{\smallskip}
 {\bf Seebeck coefficient} \qquad
 & $S_\perp$\qquad &   Eq.~\eqref{eq:Sperpgeneral}, Figs.~\ref{fig:Sperp}, \ref{fig:Sperp1} \\
  \noalign{\smallskip}
\cline{2-3}\noalign{\smallskip}
& $S_\parallel$ \qquad
&
Eq.~\eqref{eq:Sparallelgeneral}, Figs.~\ref{fig:Sperp}, \ref{fig:Sperp1}
 \\
 \noalign{\smallskip}
\cline{1-3}\noalign{\smallskip}
{\bf Nernst coeffiient}\qquad & $\eta$ & Eq.~\eqref{eq:eta}, Fig.~\ref{Figure5}\\
 \noalign{\smallskip}
\cline{1-3}\noalign{\smallskip}
{\bf Thermal conductivity}\qquad & $\kappa_\perp$ &  Eq.~\eqref{eq:kappa_perp}, Fig.~S2\\
 \noalign{\smallskip}
\cline{2-3}\noalign{\smallskip}
& $\kappa_\parallel$ \qquad
&
Eq.~\eqref{eq:kappa_parallel}, Fig.~S2\\
 \noalign{\smallskip}
\cline{1-3}\noalign{\smallskip}
{\bf Righi-Leduc coefficient}\qquad & $\mathcal{L}$ & Eq.~\eqref{eq:L}, Fig.~S3\\
\noalign{\smallskip}\hline\hline
\end{tabular}
\caption{The transport coefficients studied in this paper. The coefficients are defined through Eqs.~\eqref{eq:Efinal}, \eqref{eq:JTfinal} and calculated on the basis of the Boltzmann equation displayed in Eqs.~\eqref{LinearizedBoltzmannEq_Magnetic_0}-\eqref{Coll_ee}. The table gives the equations in which the results for the coefficients are stated and the figures in which their temperature and magnetic field dependence is illustrated. The coefficients $\rho_\parallel=\rho_\perp(B=0)$, $S_\parallel=S_\perp(B=0)$ and $\kappa_\parallel=\kappa_\perp(B=0)$ are described by the black solid lines in these figures. }
\label{table:overview}
\end{table}

Table~\ref{table:overview} provides a guide to the results obtained for the different transport coefficients in this paper and the figures that serve as illustrations. It is worth stressing several peculiarities. Only the thermal conductivities $\kappa_{\perp,\parallel}$ and the thermal Hall (Righi-Leduc) coefficient $\mathcal{L}$ depend on $1/\tau_{ee}$ even for a constant elastic scattering rate, in contrast to the electrical resistances $\rho_{\perp/\parallel}$, the Hall coefficient $R_H$, the Seebeck coefficients $S_{\perp/\parallel}$ and the Nernst coefficient $\eta$. The Hall coefficient $R_H$ displays a very weak dependence on both the electron-electron scattering rate and the momentum dependent part of the electron-impurity scattering rate $\Gamma_{\bf p}$, as long as the latter is weak compared to $1/\tau_{ei}$. The coefficients $\rho_{\perp,\parallel}$, $S_{\perp/\parallel}$ and $\mathcal{\eta}$ only depend on $1/\tau_{ee}$ if $1/\tau_{ei,{\bf p}}$ is momentum-dependent. For $S_{\perp,\parallel}$, we argued that the correction originating from a finite $\delta \Gamma_{\bf p}$ can be of the same order as the result obtained for $\delta \Gamma_{\bf p}=0$. In the case of $\eta$, a finite $\delta\Gamma_{\bf p}$ is even more impactful, since $\eta=0$ for $\delta \Gamma_{\bf p}=0$.

For all coefficients, the competition between $\tau_{ei,{\bf p}}$ and $\tau_{ee}$ plays an important role for the temperature dependence. The magnetic field enters in combination with the total scattering time as $\omega_c\tilde{\tau}_{\bf p}$. This product can contribute to the temperature dependence in two ways: first, directly through the temperature dependence of $\tau_{ee}$, and second more indirectly via the momentum dependence of $\tau_{ei,{\bf p}}$ which induces a further sensitivity of the transport coefficients to the occupation of states in momentum space. The temperature dependence of the transport coefficients becomes particularly intriguing when the energy scales $1/\tau_{ei}$, $1/\tau_{ee}$ and $\omega_c$ are of the same order. We analyzed the temperature and magnetic field dependence of the Seebeck coefficient in this case, which shows a striking qualitative similarity with experimental results on the Seebeck coefficient of Si:P on the metallic side of the $3d$ metal-insulator transition. A detailed analysis of the experimental results, however, is beyond the scope of this work. We hope that the results obtained within the simple model system studied here can serve as a guide for experimental studies of the electron kinetics at not too low temperatures.

\acknowledgments

We thank K. Michaeli for discussions. This work was supported by the College of Arts and Sciences at the University of Alabama (W.~L., G.~S.) and the National Science Foundation (NSF) under Grant No. DMR-1742752 (G.~S.) and the Army Research Office (ARO) under Grant No. W911NF2010013 and W911NF-16-1-0182 (W.~L.).

\appendix
\section{Derivation of the transport coefficients}
\label{app:derivation}

Inserting the expression for $\delta f$ stated in Eq.~\eqref{TrialSolution_Magnetic} into the defining relations for the electric and thermal currents, Eqs.~\eqref{eq:JEbasic} and Eqs.~\eqref{eq:JTbasic}, we find the following set of equations
 \begin{align}
&\mathbb{M}_0{\bf E}=\frac{m}{\mathcal{N}e^2}\left(1-\frac{1}{\tau_{ee}}\mathbb{M}_0\right){\bf J}_E-\frac{1}{eT}\mathbb{M}_1\nabla T,\label{eq:eq1}\\
&{\bf J}_T=-\frac{\mathcal{N}}{mT}\mathbb{M}_2\nabla T-\frac{\mathcal{N}e}{m}\mathbb{M}_1\left({\bf E}+\frac{m}{\mathcal{N}e^2\tau_{ee}}{\bf J}_E\right).\label{eq:eq2}
\end{align}
Here, we defined the three matrices
\begin{align}
\mathbb{M}_i=Y_{i0}+Y_{i1}(\hat{n}_{\bf B}\times)+Y_{i2}\hat{n}_{\bf B}(\hat{n}_{\bf B}\cdot), \; i\in\{0,1,2\},
\end{align}
where the matrix $Y_{mn}$ is defined in Eq.~\eqref{Ymn}.

In order to find ${\bf E}$ and ${\bf J}_T$ as functions of ${\bf J}_E$ and $\nabla T$, Eq.~\eqref{eq:eq1} can be solved for ${\bf E}$, and may be used to eliminate the electric field from the second equation in favor of ${\bf J}_E$ and $\nabla T$. These steps result in the two equations,
\begin{align}
{\bf E}&=\frac{m}{\mathcal{N} e^2 \tau_{ee}}\left(\tau_{ee}\mathbb{M}_0^{-1}-1\right){\bf J}_E-\frac{1}{eT}\mathbb{M}_0^{-1}\mathbb{M}_1\nabla T,\label{eq:eq1a}\\
{\bf J}_T&=-\frac{1}{e}\mathbb{M}_1\mathbb{M}_0^{-1}{\bf J}_E-\frac{\mathcal{N}}{mT}[\mathbb{M}_2-\mathbb{M}_1\mathbb{M}_0^{-1}\mathbb{M}_1]\nabla T.\label{eq:eq2a}
\end{align}
By comparison with Eq.~\eqref{eq:experiment}, we find
\begin{align}
\hat{\rho}&=\frac{m}{\mathcal{N} e^2 \tau_{ee}}\left(\tau_{ee}\mathbb{M}_0^{-1}-1\right),\\
\hat{\alpha}&=-\frac{1}{eT}\mathbb{M}_0^{-1}\mathbb{M}_1,\\
\hat{\pi}&=-\frac{1}{e}\mathbb{M}_1\mathbb{M}_0^{-1},\\
\hat{\kappa}&=\frac{\mathcal{N}}{mT}[\mathbb{M}_2-\mathbb{M}_1\mathbb{M}_0^{-1}\mathbb{M}_1].
\end{align}
In order to find explicit expressions for these tensors, we need to know the inverse of the matrix $\mathbb{M}_0$,
\begin{align}
\mathbb{M}_0^{-1}=\frac{Y_{00}-Y_{01}(\hat{n}_{\bf B}\times)}{Y_{00}^2+Y_{01}^2}+\frac{(Y_{01}^2-Y_{00}Y_{02})\hat{n}_{\bf B}(\hat{n}_{\bf B}\cdot)}{(Y_{00}^2+Y_{01}^2)(Y_{00}+Y_{02})}.
\end{align}
It is easily checked that the two matrices $\mathbb{M}_0^{-1}$ and $\mathbb{M}_1$ commute, so that $\hat{\pi}= T\hat{\alpha}$. The relations stated above enable us to find the transport coefficients given in Eqs.~\eqref{eq:rhoperp}-\eqref{eq:rhoparallel},~\eqref{eq:Sperpgeneral}-\eqref{eq:kappa_parallel}. Let us note that the expression for $\rho_\parallel$ was obtained using the relation $Y_{m0}+Y_{m2}=\left\langle\!\left\langle \xi_{\bf p}^m\tilde{\tau}_{\bf p}\right\rangle\!\right\rangle$ for $m=0$. In particular, the combination $ Y_{m0}+Y_{m2}$ is magnetic field-independent, and so is $\rho_\parallel=\rho_\perp(B=0)$.

\renewcommand{\thefigure}{S\arabic{figure}}

\section*{Supplementary Material}
\setcounter{figure}{0}

In this supplementary material, we discuss the temperature and magnetic field dependence of the Hall coefficient $R_H$, the thermal conductivity $\kappa_\perp$, and thermal Hall coefficient $\mathcal{L}$.

\subsection{Hall coefficient $R_H$}
\label{subsec:RH}
\begin{figure}[t]
\centering
\includegraphics[width=0.45\textwidth]
{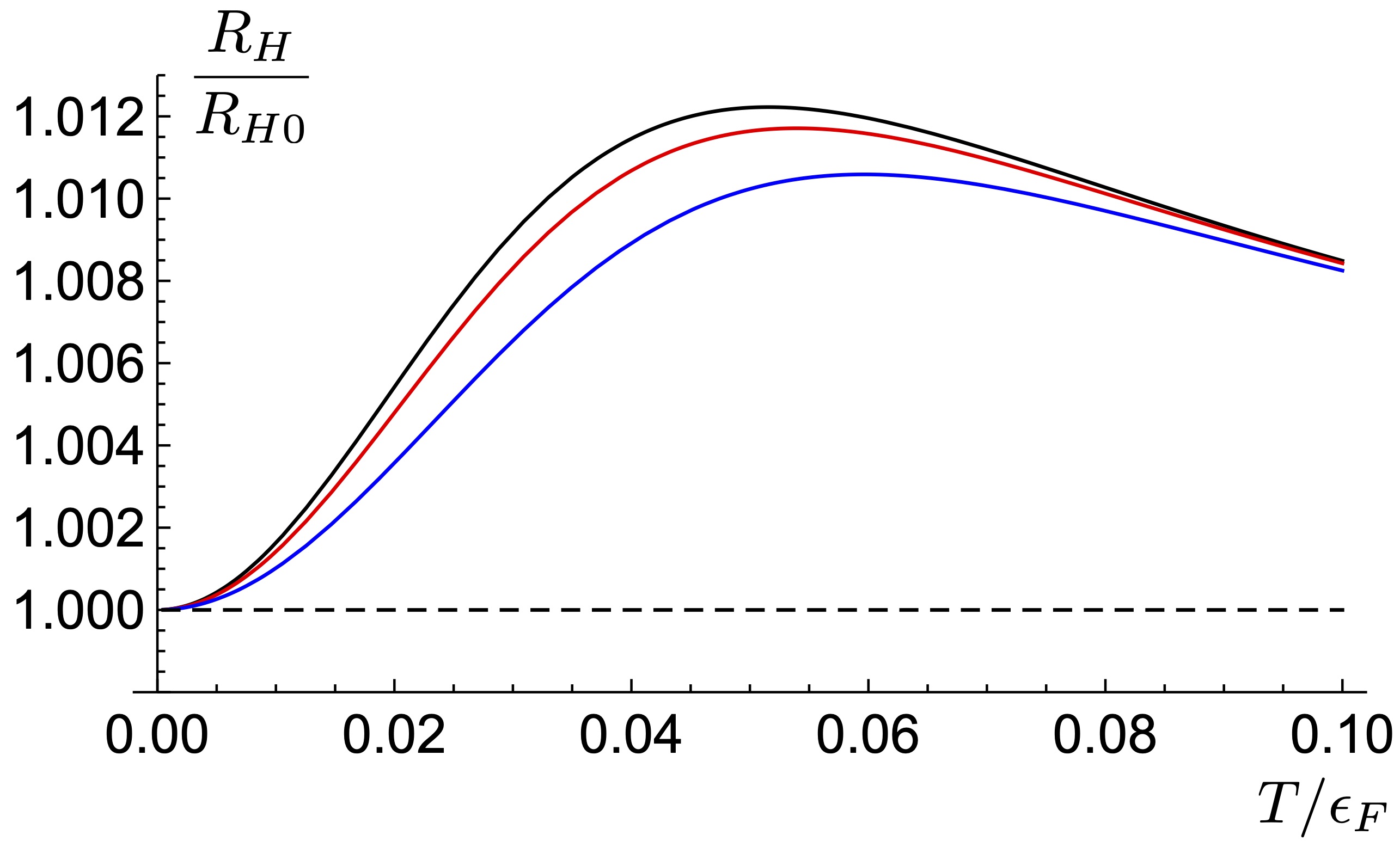} \\
\caption{
The Hall coefficient $R_H$, Eq.~(23), normalized to $R_{H0}$, Eq.~(47), as a function of temperature $T$, and for different magnetic fields. Parameters as listed in the caption of Fig.~2. 
}
\label{fig:RH}
\end{figure}

The Hall coefficient $R_H$ is temperature and magnetic field-independent for $\delta \Gamma_{\bf p}=0$,
\begin{align}
R_H\rightarrow R_{H0}=-\frac{1}{\mathcal{N} e}, \qquad \tau_{ei}=\mbox{const.}
\end{align}
For a finite $\delta \Gamma_{\bf p}$, the Hall coefficient goes through a maximum at finite temperatures, and also displays a weak magnetic field dependence, see Fig.~\ref{fig:RH}. The overall magnitude of the corrections to $R_{H0}$, however, is small. Formally, this remarkable insensitivity to $\Gamma_{\bf p}$ is readily understood by noting that the first order correction in $\delta \Gamma_{\bf p}$ vanishes, $\delta R^{(1)}_H=0$.

\subsection{Thermal conductivity $\kappa_\perp$}
\label{subsec:kappa_perp}

The expression for the thermal conductivity for a constant elastic scattering rate, $\kappa_\perp=\kappa_0$, is given in Eq.~(47). We will first focus our discussion on the case $B=0$, for which $\kappa_0\propto T\tilde{\tau}$ holds at low temperatures $T\ll \epsilon_F$. In this regime, the temperature dependence is determined by the competition of the factor $T$ with the temperature dependence of $\tilde{\tau}(T)$. The first factor, $T$, dominates at the lowest temperatures, where the condition $\tau_{ee}>\tau_{ei}$ is fulfilled and $\tilde{\tau}$ is approximately constant. When increasing temperature, $\tilde{\tau}$ starts decreasing appreciably as soon as $\tau_{ee}<\tau_{ei}$. The overall result is a maximum in $\kappa_0$ at finite temperatures. For finite magnetic fields, $\kappa_0$ decreases due to the factor $[1+\omega_c^2\tilde{\tau}^2]^{-1}$. This effect diminishes with increasing temperature, when $\tilde{\tau}$ becomes shorter.

\begin{figure}[tbhp]
\centering
\includegraphics[width=0.45\textwidth]
{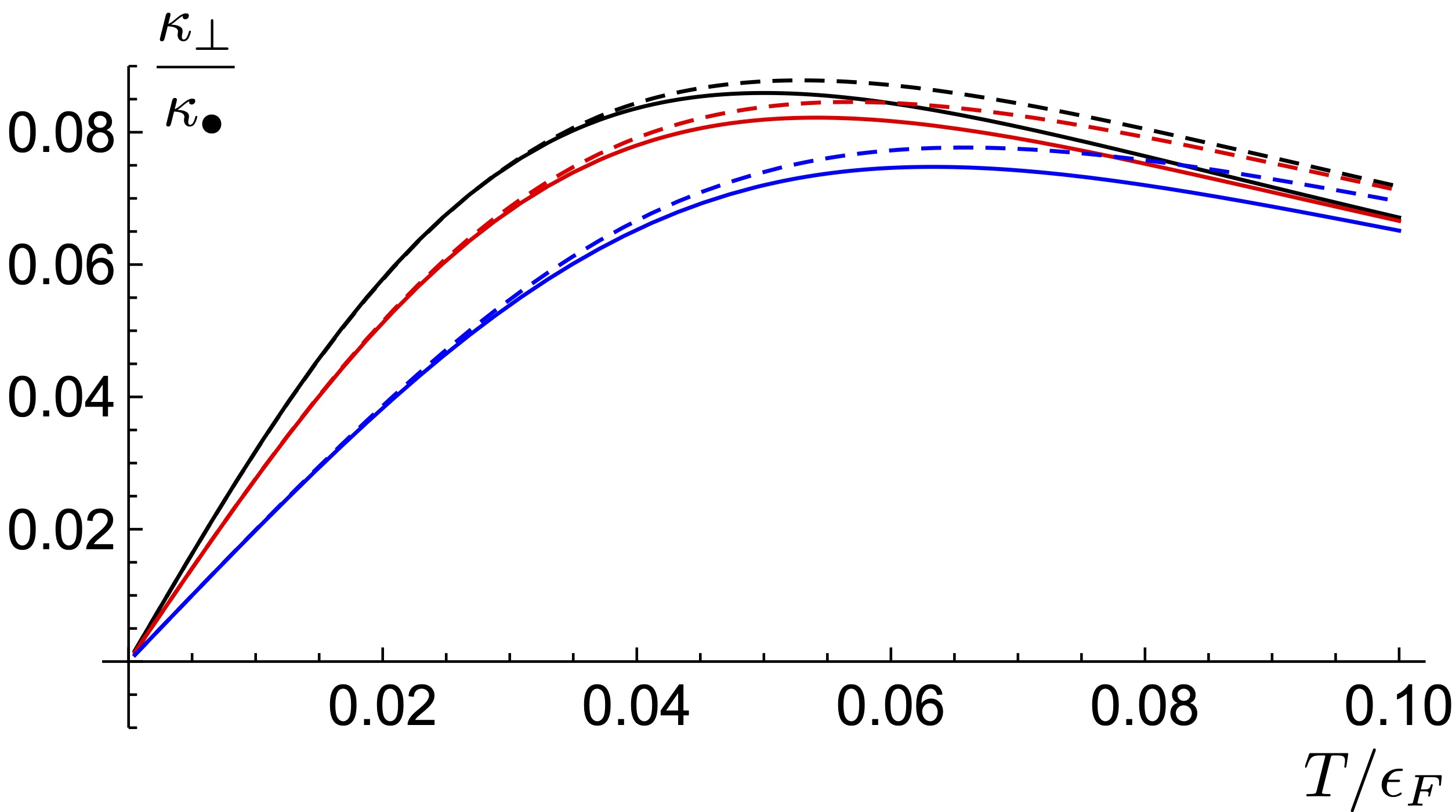} \\
\caption{The thermal conductivity $\kappa_\perp$, Eq.~(30), normalized to $\kappa_\bullet=\mathcal{N}\epsilon_F\tau_{ei}/m$, as a function of temperature $T$, and for different magnetic fields. In particular, the black line stands for $\kappa_\perp(B=0)=\rho_\parallel$. Parameters as listed in the caption of Fig.~2. }
\label{Figure6}
\end{figure}

Fig.~\ref{Figure6} illustrates the temperature and magnetic field dependence of $\kappa_0$ (dashed lines) and confirms the general trends discussed above. This figure also shows the results for finite $\delta\Gamma_{\bf p}$ (solid lines). Particle-hole asymmetry is not essential for the leading contribution to the thermal conductivity $\kappa_\perp=\kappa_0$, which is proportional to $\left\langle\!\left\langle \xi_{\bf p}^2\right\rangle\!\right\rangle$. Correspondingly, one expects the corrections due to finite $\delta\Gamma_{\bf p}$ to be small in the parameter $T/\epsilon_F$. The smallness of the correction $\kappa_\perp-\kappa_0$ compared to $\kappa_0$ is indeed visible in Fig.~\ref{Figure6}.

\subsection{Thermal Hall (Righi-Leduc) coefficient $\mathcal{L}$}
\label{subsec:L}

\begin{figure}[tbhp]
\centering
\includegraphics[width=0.45\textwidth]
{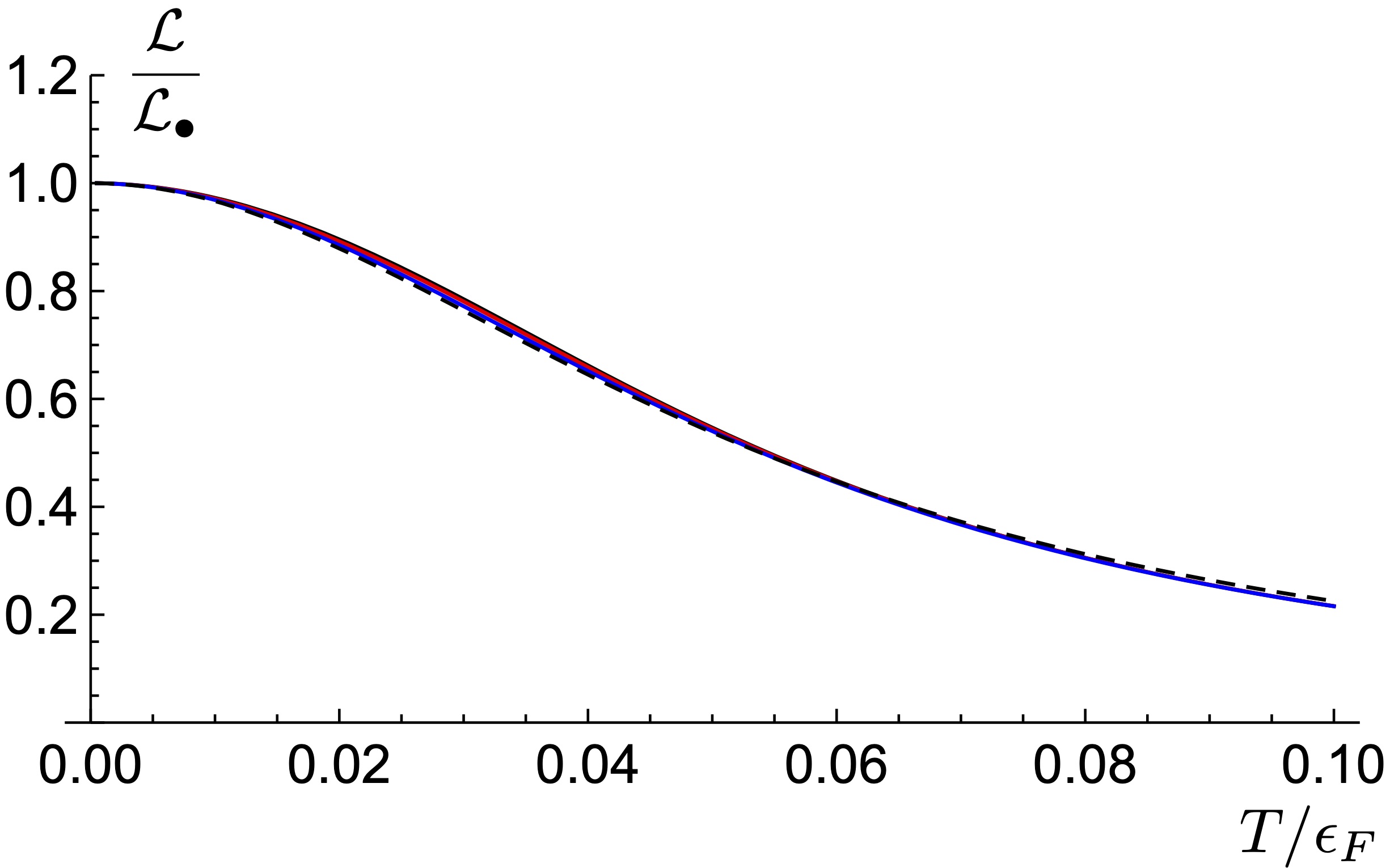} \\
\caption{Thermal Hall (Righi-Leduc) coefficient $\mathcal{L}$, Eq.~(31), normalized to $\mathcal{L}_\bullet=-e\tau_{ei}/m$, as a function of temperature $T$, and for different magnetic fields. Parameters as listed in the caption of Fig.~2. 
}
\label{Figure7}
\end{figure}

For a constant elastic scattering rate, $\mathcal{L}=\mathcal{L}_0=-e\tilde{\tau}/m$ is temperature dependent, but not magnetic field dependent. The temperature dependence is inherited from the inelastic scattering rate only. As in the case of thermal conductivity, particle-hole asymmetry does not affect appreciably the leading contribution $\mathcal{L}_0$ obtained for $\delta \Gamma_{\bf p}=0$, and the corrections due to finite $\delta\Gamma$ are expected to be small in $T/\epsilon_F$ in comparison. We see in Fig.~\ref{Figure7} that the deviation of $\mathcal{L}$ (solid lines) from $\mathcal{L}_0$ (dashed line) is indeed small, if we plot the results for the same parameters used for the other transport coefficients. The Righi-Leduc coefficient $\mathcal{L}$ is well approximated by $\mathcal{L}_0=-e\tilde{\tau}/m$ at low temperatures, even for finite magnetic fields. While a magnetic field affects $\mathcal{L}$ only weakly, a closer inspection of the data plotted in Fig.~\ref{Figure7} reveals that it tends to suppress the influence of a finite $\delta\Gamma_{\bf p}$, as already observed for the other coefficients.

\begin{thebibliography}{28}
\expandafter\ifx\csname natexlab\endcsname\relax\def\natexlab#1{#1}\fi
\expandafter\ifx\csname bibnamefont\endcsname\relax
  \def\bibnamefont#1{#1}\fi
\expandafter\ifx\csname bibfnamefont\endcsname\relax
  \def\bibfnamefont#1{#1}\fi
\expandafter\ifx\csname citenamefont\endcsname\relax
  \def\citenamefont#1{#1}\fi
\expandafter\ifx\csname url\endcsname\relax
  \def\url#1{\texttt{#1}}\fi
\expandafter\ifx\csname urlprefix\endcsname\relax\def\urlprefix{URL }\fi
\providecommand{\bibinfo}[2]{#2}
\providecommand{\eprint}[2][]{\url{#2}}

\bibitem[{\citenamefont{Tsuji}(1958)}]{Tsuji58}
\bibinfo{author}{\bibfnamefont{M.}~\bibnamefont{Tsuji}},
  \bibinfo{journal}{Journal of the Physical Society of Japan}
  \textbf{\bibinfo{volume}{13}}, \bibinfo{pages}{979} (\bibinfo{year}{1958}).

\bibitem[{\citenamefont{Beer}(1963)}]{Beer63}
\bibinfo{author}{\bibfnamefont{A.}~\bibnamefont{Beer}},
  \emph{\bibinfo{title}{Galvanomagnetic effects in semiconductors}}
  (\bibinfo{publisher}{Academic Press (New York and London)},
  \bibinfo{year}{1963}).

\bibitem[{\citenamefont{Ziman}(2001)}]{Ziman01}
\bibinfo{author}{\bibfnamefont{J.~M.} \bibnamefont{Ziman}},
  \emph{\bibinfo{title}{Electrons and Phonons}} (\bibinfo{publisher}{Oxford
  University Press}, \bibinfo{year}{2001}).

\bibitem[{\citenamefont{Gantmakher and Levinson}(2012)}]{Gantmakher12}
\bibinfo{author}{\bibfnamefont{V.~F.} \bibnamefont{Gantmakher}}
  \bibnamefont{and} \bibinfo{author}{\bibfnamefont{Y.~B.}
  \bibnamefont{Levinson}}, \emph{\bibinfo{title}{Carrier scattering in metals
  and semiconductors}} (\bibinfo{publisher}{North Holland},
  \bibinfo{year}{2012}).

\bibitem[{\citenamefont{Gurzhi}(1963)}]{Gurzhi63}
\bibinfo{author}{\bibfnamefont{R.~N.} \bibnamefont{Gurzhi}},
  \bibinfo{journal}{ZETF} \textbf{\bibinfo{volume}{44}}, \bibinfo{pages}{771}
  (\bibinfo{year}{1963}), \bibinfo{note}{[{S}ov. Phys. JETP {\bf 17}, 521
  (1963)]}.

\bibitem[{\citenamefont{de~Jong and Molenkamp}(1995)}]{deJong95}
\bibinfo{author}{\bibfnamefont{M.~J.~M.} \bibnamefont{de~Jong}}
  \bibnamefont{and} \bibinfo{author}{\bibfnamefont{L.~W.}
  \bibnamefont{Molenkamp}}, \bibinfo{journal}{Phys. Rev. B}
  \textbf{\bibinfo{volume}{51}}, \bibinfo{pages}{13389} (\bibinfo{year}{1995}).

\bibitem[{\citenamefont{Andreev et~al.}(2011)\citenamefont{Andreev, Kivelson,
  and Spivak}}]{Andreev11}
\bibinfo{author}{\bibfnamefont{A.~V.} \bibnamefont{Andreev}},
  \bibinfo{author}{\bibfnamefont{S.~A.} \bibnamefont{Kivelson}},
  \bibnamefont{and} \bibinfo{author}{\bibfnamefont{B.}~\bibnamefont{Spivak}},
  \bibinfo{journal}{Phys. Rev. Lett.} \textbf{\bibinfo{volume}{106}},
  \bibinfo{pages}{256804} (\bibinfo{year}{2011}).

\bibitem[{\citenamefont{Mahajan et~al.}(2013)\citenamefont{Mahajan, Barkeshli,
  and Hartnoll}}]{Mahajan13}
\bibinfo{author}{\bibfnamefont{R.}~\bibnamefont{Mahajan}},
  \bibinfo{author}{\bibfnamefont{M.}~\bibnamefont{Barkeshli}},
  \bibnamefont{and} \bibinfo{author}{\bibfnamefont{S.~A.}
  \bibnamefont{Hartnoll}}, \bibinfo{journal}{Phys. Rev. B}
  \textbf{\bibinfo{volume}{88}}, \bibinfo{pages}{125107}
  (\bibinfo{year}{2013}).

\bibitem[{\citenamefont{Xie and Foster}(2016)}]{Xie16}
\bibinfo{author}{\bibfnamefont{H.-Y.} \bibnamefont{Xie}} \bibnamefont{and}
  \bibinfo{author}{\bibfnamefont{M.~S.} \bibnamefont{Foster}},
  \bibinfo{journal}{Phys. Rev. B} \textbf{\bibinfo{volume}{93}},
  \bibinfo{pages}{195103} (\bibinfo{year}{2016}).

\bibitem[{\citenamefont{Moll et~al.}(2016)\citenamefont{Moll, Kushwaha, Nandi,
  Schmidt, and Mackenzie}}]{Moll16}
\bibinfo{author}{\bibfnamefont{P.~J.~W.} \bibnamefont{Moll}},
  \bibinfo{author}{\bibfnamefont{P.}~\bibnamefont{Kushwaha}},
  \bibinfo{author}{\bibfnamefont{N.}~\bibnamefont{Nandi}},
  \bibinfo{author}{\bibfnamefont{B.}~\bibnamefont{Schmidt}}, \bibnamefont{and}
  \bibinfo{author}{\bibfnamefont{A.~P.} \bibnamefont{Mackenzie}},
  \textbf{\bibinfo{volume}{351}}, \bibinfo{pages}{1061} (\bibinfo{year}{2016}).

\bibitem[{\citenamefont{Crossno et~al.}(2016)\citenamefont{Crossno, Shi, Wang,
  Liu, Harzheim, Lucas, Sachdev, Kim, Taniguchi, Watanabe et~al.}}]{Crossno16}
\bibinfo{author}{\bibfnamefont{J.}~\bibnamefont{Crossno}},
  \bibinfo{author}{\bibfnamefont{J.~K.} \bibnamefont{Shi}},
  \bibinfo{author}{\bibfnamefont{K.}~\bibnamefont{Wang}},
  \bibinfo{author}{\bibfnamefont{X.}~\bibnamefont{Liu}},
  \bibinfo{author}{\bibfnamefont{A.}~\bibnamefont{Harzheim}},
  \bibinfo{author}{\bibfnamefont{A.}~\bibnamefont{Lucas}},
  \bibinfo{author}{\bibfnamefont{S.}~\bibnamefont{Sachdev}},
  \bibinfo{author}{\bibfnamefont{P.}~\bibnamefont{Kim}},
  \bibinfo{author}{\bibfnamefont{T.}~\bibnamefont{Taniguchi}},
  \bibinfo{author}{\bibfnamefont{K.}~\bibnamefont{Watanabe}},
  \bibnamefont{et~al.}, \textbf{\bibinfo{volume}{351}}, \bibinfo{pages}{1058}
  (\bibinfo{year}{2016}).

\bibitem[{\citenamefont{Bandurin et~al.}(2016)\citenamefont{Bandurin, Torre,
  Kumar, Ben~Shalom, Tomadin, Principi, Auton, Khestanova, Novoselov,
  Grigorieva et~al.}}]{Bandurin16}
\bibinfo{author}{\bibfnamefont{D.~A.} \bibnamefont{Bandurin}},
  \bibinfo{author}{\bibfnamefont{I.}~\bibnamefont{Torre}},
  \bibinfo{author}{\bibfnamefont{R.~K.} \bibnamefont{Kumar}},
  \bibinfo{author}{\bibfnamefont{M.}~\bibnamefont{Ben~Shalom}},
  \bibinfo{author}{\bibfnamefont{A.}~\bibnamefont{Tomadin}},
  \bibinfo{author}{\bibfnamefont{A.}~\bibnamefont{Principi}},
  \bibinfo{author}{\bibfnamefont{G.~H.} \bibnamefont{Auton}},
  \bibinfo{author}{\bibfnamefont{E.}~\bibnamefont{Khestanova}},
  \bibinfo{author}{\bibfnamefont{K.~S.} \bibnamefont{Novoselov}},
  \bibinfo{author}{\bibfnamefont{I.~V.} \bibnamefont{Grigorieva}},
  \bibnamefont{et~al.}, \textbf{\bibinfo{volume}{351}}, \bibinfo{pages}{1055}
  (\bibinfo{year}{2016}).

\bibitem[{\citenamefont{Narozhny et~al.}(2017)\citenamefont{Narozhny, Gornyi,
  Mirlin, and Schmalian}}]{Narozhny17}
\bibinfo{author}{\bibfnamefont{B.~N.} \bibnamefont{Narozhny}},
  \bibinfo{author}{\bibfnamefont{I.~V.} \bibnamefont{Gornyi}},
  \bibinfo{author}{\bibfnamefont{A.~D.} \bibnamefont{Mirlin}},
  \bibnamefont{and}
  \bibinfo{author}{\bibfnamefont{J.}~\bibnamefont{Schmalian}},
  \bibinfo{journal}{Annalen der Physik} \textbf{\bibinfo{volume}{529}},
  \bibinfo{pages}{1700043} (\bibinfo{year}{2017}).

\bibitem[{\citenamefont{Guo et~al.}(2017)\citenamefont{Guo, Ilseven, Falkovich,
  and Levitov}}]{Guo17}
\bibinfo{author}{\bibfnamefont{H.}~\bibnamefont{Guo}},
  \bibinfo{author}{\bibfnamefont{E.}~\bibnamefont{Ilseven}},
  \bibinfo{author}{\bibfnamefont{G.}~\bibnamefont{Falkovich}},
  \bibnamefont{and} \bibinfo{author}{\bibfnamefont{L.~S.}
  \bibnamefont{Levitov}}, \bibinfo{journal}{Proceedings of the National Academy
  of Sciences} \textbf{\bibinfo{volume}{114}}, \bibinfo{pages}{3068}
  (\bibinfo{year}{2017}).

\bibitem[{\citenamefont{Levchenko et~al.}(2017)\citenamefont{Levchenko, Xie,
  and Andreev}}]{Levchenko17}
\bibinfo{author}{\bibfnamefont{A.}~\bibnamefont{Levchenko}},
  \bibinfo{author}{\bibfnamefont{H.-Y.} \bibnamefont{Xie}}, \bibnamefont{and}
  \bibinfo{author}{\bibfnamefont{A.~V.} \bibnamefont{Andreev}},
  \bibinfo{journal}{Phys. Rev. B} \textbf{\bibinfo{volume}{95}},
  \bibinfo{pages}{121301} (\bibinfo{year}{2017}).

\bibitem[{\citenamefont{Lucas and Das~Sarma}(2018)}]{Lucas18}
\bibinfo{author}{\bibfnamefont{A.}~\bibnamefont{Lucas}} \bibnamefont{and}
  \bibinfo{author}{\bibfnamefont{S.}~\bibnamefont{Das~Sarma}},
  \bibinfo{journal}{Phys. Rev. B} \textbf{\bibinfo{volume}{97}},
  \bibinfo{pages}{245128} (\bibinfo{year}{2018}).

\bibitem[{\citenamefont{Zarenia et~al.}(2019)\citenamefont{Zarenia, Principi,
  and Vignale}}]{Zarenia19}
\bibinfo{author}{\bibfnamefont{M.}~\bibnamefont{Zarenia}},
  \bibinfo{author}{\bibfnamefont{A.}~\bibnamefont{Principi}}, \bibnamefont{and}
  \bibinfo{author}{\bibfnamefont{G.}~\bibnamefont{Vignale}},
  \bibinfo{journal}{2D Materials} \textbf{\bibinfo{volume}{6}},
  \bibinfo{pages}{035024} (\bibinfo{year}{2019}).

\bibitem[{\citenamefont{Principi and Vignale}(2015)}]{Principi15}
\bibinfo{author}{\bibfnamefont{A.}~\bibnamefont{Principi}} \bibnamefont{and}
  \bibinfo{author}{\bibfnamefont{G.}~\bibnamefont{Vignale}},
  \bibinfo{journal}{Phys. Rev. Lett.} \textbf{\bibinfo{volume}{115}},
  \bibinfo{pages}{056603} (\bibinfo{year}{2015}).

\bibitem[{\citenamefont{Lee et~al.}(2020)\citenamefont{Lee, Finkel'stein,
  Michaeli, and Schwiete}}]{Lee20}
\bibinfo{author}{\bibfnamefont{W.-R.} \bibnamefont{Lee}},
  \bibinfo{author}{\bibfnamefont{A.~M.} \bibnamefont{Finkel'stein}},
  \bibinfo{author}{\bibfnamefont{K.}~\bibnamefont{Michaeli}}, \bibnamefont{and}
  \bibinfo{author}{\bibfnamefont{G.}~\bibnamefont{Schwiete}},
  \bibinfo{journal}{Phys. Rev. Research} \textbf{\bibinfo{volume}{2}},
  \bibinfo{pages}{013148} (\bibinfo{year}{2020}).

\bibitem[{\citenamefont{Keyes}(1958)}]{Keyes58}
\bibinfo{author}{\bibfnamefont{R.~W.} \bibnamefont{Keyes}},
  \bibinfo{journal}{Journal of Physics and Chemistry of Solids}
  \textbf{\bibinfo{volume}{6}}, \bibinfo{pages}{1 } (\bibinfo{year}{1958}).

\bibitem[{\citenamefont{Lakner and L\"ohneysen}(1993)}]{Lakner93}
\bibinfo{author}{\bibfnamefont{M.}~\bibnamefont{Lakner}} \bibnamefont{and}
  \bibinfo{author}{\bibfnamefont{H.~v.} \bibnamefont{L\"ohneysen}},
  \bibinfo{journal}{Phys. Rev. Lett.} \textbf{\bibinfo{volume}{70}},
  \bibinfo{pages}{3475} (\bibinfo{year}{1993}).

\bibitem[{Sup()}]{Suppl}
\bibinfo{note}{See Supplementary Material for the discussion and
  illustration of the temperature and magnetic field dependence of the
  transport coefficients $R_H$, $\kappa_\perp$,
  $\kappa_\parallel=\kappa_\perp(B=0)$, and $\mathcal{L}$.}

\bibitem[{Ele()}]{Electrochem}
\bibinfo{note}{For the sake of simplicity, we do not distinguish between ${\bf
  E}=-\nabla \varphi$ and ${\bf E}'=\frac{1}{e}\nabla (\mu-e\varphi)$, where
  $-e$ is the charge of the electron, $\varphi$ the electric potential and
  $\mu$ the chemical potential}.

\bibitem[{\citenamefont{Onsager}(1931{\natexlab{a}})}]{Onsager31}
\bibinfo{author}{\bibfnamefont{L.}~\bibnamefont{Onsager}},
  \bibinfo{journal}{Phys. Rev.} \textbf{\bibinfo{volume}{37}},
  \bibinfo{pages}{405} (\bibinfo{year}{1931}{\natexlab{a}}).

\bibitem[{\citenamefont{Onsager}(1931{\natexlab{b}})}]{Onsager31a}
\bibinfo{author}{\bibfnamefont{L.}~\bibnamefont{Onsager}},
  \bibinfo{journal}{Phys. Rev.} \textbf{\bibinfo{volume}{38}},
  \bibinfo{pages}{2265} (\bibinfo{year}{1931}{\natexlab{b}}).

\bibitem[{\citenamefont{van Vliet}(2008)}]{vanVliet08}
\bibinfo{author}{\bibfnamefont{C.~M.} \bibnamefont{van Vliet}},
  \emph{\bibinfo{title}{Equilibrium and Non-Equilibrium Statistical Mechanics}}
  (\bibinfo{publisher}{World Scientific}, \bibinfo{year}{2008}).

\bibitem[{\citenamefont{Levinson}(1977)}]{Levinson77}
\bibinfo{author}{\bibfnamefont{I.~B.} \bibnamefont{Levinson}},
  \bibinfo{journal}{Zh. Eksp. Teor. Fiz.} \textbf{\bibinfo{volume}{73}},
  \bibinfo{pages}{318} (\bibinfo{year}{1977}), \bibinfo{note}{[Sov. Phys. JETP
  {\bf 46}, 165 (1977)]}.

\bibitem[{\citenamefont{Maki}(1969)}]{Maki69}
\bibinfo{author}{\bibfnamefont{K.}~\bibnamefont{Maki}},
  \bibinfo{journal}{Progress of Theoretical Physics}
  \textbf{\bibinfo{volume}{41}}, \bibinfo{pages}{586} (\bibinfo{year}{1969}).

\end{thebibliography}

\end{document}